\PassOptionsToPackage{unicode}{hyperref}
\PassOptionsToPackage{hyphens}{url}
\PassOptionsToPackage{dvipsnames,svgnames,x11names}{xcolor}
\documentclass[
  12pt]{article}
\usepackage{xcolor}
\usepackage{amsmath,amssymb}
\usepackage{iftex}
\ifPDFTeX
  \usepackage[T1]{fontenc}
  \usepackage[utf8]{inputenc}
  \usepackage{textcomp} % provide euro and other symbols
\else % if luatex or xetex
  \usepackage{unicode-math} % this also loads fontspec
  \defaultfontfeatures{Scale=MatchLowercase}
  \defaultfontfeatures[\rmfamily]{Ligatures=TeX,Scale=1}
\fi
\usepackage{lmodern}
\ifPDFTeX\else
\fi
\IfFileExists{upquote.sty}{\usepackage{upquote}}{}
\IfFileExists{microtype.sty}{% use microtype if available
  \usepackage[]{microtype}
  \UseMicrotypeSet[protrusion]{basicmath} % disable protrusion for tt fonts
}{}
\makeatletter
\@ifundefined{KOMAClassName}{% if non-KOMA class
  \IfFileExists{parskip.sty}{%
    \usepackage{parskip}
  }{% else
    \setlength{\parindent}{0pt}
    \setlength{\parskip}{6pt plus 2pt minus 1pt}}
}{% if KOMA class
  \KOMAoptions{parskip=half}}
\makeatother
\makeatletter
\ifx\paragraph\undefined\else
  \let\oldparagraph\paragraph
  \renewcommand{\paragraph}{
    \@ifstar
      \xxxParagraphStar
      \xxxParagraphNoStar
  }
  \newcommand{\xxxParagraphStar}[1]{\oldparagraph*{#1}\mbox{}}
  \newcommand{\xxxParagraphNoStar}[1]{\oldparagraph{#1}\mbox{}}
\fi
\ifx\subparagraph\undefined\else
  \let\oldsubparagraph\subparagraph
  \renewcommand{\subparagraph}{
    \@ifstar
      \xxxSubParagraphStar
      \xxxSubParagraphNoStar
  }
  \newcommand{\xxxSubParagraphStar}[1]{\oldsubparagraph*{#1}\mbox{}}
  \newcommand{\xxxSubParagraphNoStar}[1]{\oldsubparagraph{#1}\mbox{}}
\fi
\makeatother

\usepackage{longtable,booktabs,array}
\usepackage{calc} % for calculating minipage widths
\usepackage{etoolbox}
\makeatletter
\patchcmd\longtable{\par}{\if@noskipsec\mbox{}\fi\par}{}{}
\makeatother
\IfFileExists{footnotehyper.sty}{\usepackage{footnotehyper}}{\usepackage{footnote}}
\makesavenoteenv{longtable}
\usepackage{graphicx}
\makeatletter
\newsavebox\pandoc@box
\newcommand*\pandocbounded[1]{% scales image to fit in text height/width
  \sbox\pandoc@box{#1}%
  \Gscale@div\@tempa{\textheight}{\dimexpr\ht\pandoc@box+\dp\pandoc@box\relax}%
  \Gscale@div\@tempb{\linewidth}{\wd\pandoc@box}%
  \ifdim\@tempb\p@<\@tempa\p@\let\@tempa\@tempb\fi% select the smaller of both
  \ifdim\@tempa\p@<\p@\scalebox{\@tempa}{\usebox\pandoc@box}%
  \else\usebox{\pandoc@box}%
  \fi%
}
\def\fps@figure{htbp}
\makeatother

\providecommand{\tightlist}{%
  \setlength{\itemsep}{0pt}\setlength{\parskip}{0pt}}

\usepackage[]{natbib}
\usepackage{soul}
\usepackage{xcolor}

\makeatletter
\@ifpackageloaded{caption}{}{\usepackage{caption}}
\AtBeginDocument{%
\ifdefined\contentsname
  \renewcommand*\contentsname{Table of contents}
\else
  \newcommand\contentsname{Table of contents}
\fi
\ifdefined\listfigurename
  \renewcommand*\listfigurename{List of Figures}
\else
  \newcommand\listfigurename{List of Figures}
\fi
\ifdefined\listtablename
  \renewcommand*\listtablename{List of Tables}
\else
  \newcommand\listtablename{List of Tables}
\fi
\ifdefined\figurename
  \renewcommand*\figurename{Figure}
\else
  \newcommand\figurename{Figure}
\fi
\ifdefined\tablename
  \renewcommand*\tablename{Table}
\else
  \newcommand\tablename{Table}
\fi
}
\@ifpackageloaded{float}{}{\usepackage{float}}
\floatstyle{ruled}
\@ifundefined{c@chapter}{\newfloat{codelisting}{h}{lop}}{\newfloat{codelisting}{h}{lop}[chapter]}
\floatname{codelisting}{Listing}

\makeatother
\makeatletter
\@ifpackageloaded{caption}{}{\usepackage{caption}}
\@ifpackageloaded{subcaption}{}{\usepackage{subcaption}}
\makeatother
\usepackage{bookmark}
\IfFileExists{xurl.sty}{\usepackage{xurl}}{} % add URL line breaks if available
\hypersetup{
  pdftitle={Robust Indicators of Spatial Association},
  pdfauthor={Levi John Wolf; Wei Kang},
  pdfkeywords={spatial autocorrelation, robust statistics, LISA, trimmed
least squares, spatial analysis, conditional permutation},
  colorlinks=true,
  linkcolor={blue},
  filecolor={Maroon},
  citecolor={Blue},
  urlcolor={Blue},
  pdfcreator={LaTeX via pandoc}}

\begin{document}

\def\spacingset#1{\renewcommand{\baselinestretch}%
{#1}\small\normalsize} \spacingset{1}

%%%%%%%%%%%%%%%%%%%%%%%%%%%%%%%%%%%%%%%%%%%%%%%%%%%%%%%%%%%%%%%%%%%%%%%%%%%%%%

\date{July 9, 2026}
\title{\bf Robust Indicators of Spatial Association}
\author{
Levi John Wolf\\
School of Geographical Sciences, University of Bristol\\
and\\Wei Kang\\
Center for Geospatial Sciences, \\School of Public Policy, University of
California, Riverside\\
}
\maketitle

\bigskip
\bigskip
\begin{abstract}
The Moran statistic and its accompanying local statistics are among the
most widely used exploratory tools for assessing global and local
spatial autocorrelation. Their paired visualizations, the Moran
Scatterplot and LISA map, are similarly central to spatial analysis,
together helping identify spatial clusters, regions where observations
resemble their surroundings, and spatial outliers, observations that
differ sharply from them. However, using Moran statistics to detect
spatial outliers is complicated by their high sensitivity to
\emph{distributional} outliers: observations that are extreme relative
to the overall data distribution, regardless of spatial context. A
single distributional outlier can (I) distort local statistics across
the entire map and (II) bias the global estimate of spatial association.
In this paper, we offer the first systematic evaluation of robust LISA
and global spatial association measures, using a variety of plug-in
robust estimators, a trimmed least squares (TLS) estimator, and a
Theil-Sen-style estimator. We also outline a visualization strategy for
constructing Robust Moran Scatterplots and LISA maps for each estimator.
Across all approaches considered, we find that the Theil-Sen Moran
estimator is the better default for exploratory spatial data analysis
and visualization, while robust plug-in estimators offer acceptable
performance in large datasets.
\end{abstract}

\noindent%
\vfill

\newpage
\spacingset{1.9} % DON'T change the spacing!

\section{Introduction}\label{introduction}

The local Moran statistic \citep{anselin1995local} is, by a wide margin,
the most heavily used local indicator of spatial association in applied
work. Its appeal comes from two core reasons. First, it is a
decomposition of the global Moran's \(I\)
\citep{moran1948interpretation, cliff1981spatial} classifying
observations into either spatial clusters (sites that are strongly
similar to their surroundings) or spatial outliers (sites that
significantly different from their surroundings). Other common
exploratory spatial statistics, like the Getis-Ord \(G\)
\citep{getis1992analysis}, have local statistics that classify
observations into high or low clusters only \citep{ord1995local}. This
limits their usefulness in identifying anomalous observations in a map.
The second advantage for the Moran statistic is its natural paired
visualization: the Moran scatterplot offers a simple way to visualize
the local and global spatial association in a given dataset
\citep{anselin1996moran}. A paired Local Moran map and Moran Scatterplot
frequently provide the first point of entry for general statisticians or
non-geographers to assess spatial structure during exploratory spatial
data analysis.

Many studies have sought to extend or adapt the basic ideas of local
indicators of spatial association
\citep{getis1996local, lee2001developing, li2007moran, ord2012local, rey2016spacetime, anselin2018local, naimi2019elsa, wolf2024confounded}.
For much of this work on exploratory spatial statistics, innovation
occurs in moving ``beyond Moran's \(I\)'' \citep{li2007moran} to focus
on new estimands that summarise different kinds of outcomes
\citep{berglund1999identifying, li2007moran, ord2012local, rey2016spacetime, tao2016spatial, naimi2019elsa}
or generalize univariate statistics to more dimensions
\citep{dray2011revisiting, anselin2018local, wolf2024confounded}. Most
new global statistics also seek to identify a natural paired
visualization as a scatterplot, as well as the ability to classify
observations in the map into ``clusters'' or ``outliers'' relative to
their surroundings \citep{ord2023art}.

Sometimes, arguments about these local indicators critique the estimand
that different estimators or inference methods seek to recover. For
instance, \citet{lee2001developing}'s analysis of the bivariate global
association statistic by \citet{wartenberg1985multivariate} suggests an
alternative spatial correlation framework to study bivariate
association, while \citet{wolf2024confounded} goes further still to
define a generalized multivariate spatial association estimator using
partial dependence \citep{waugh1933partial}. \citet{li2007moran}
provides a critique of Moran statistics by changing estimand entirely
from the Moran-form regression to a spatial lag model
\citep{anselin1988methods}, offering both univariate global and local
statistics in this new reverse regression. Additional work has focused
on improving statistical inference methods, especially for local
indicators of spatial association
\citep[\citet{sauer2021importance}]{sokal1993testing, tiefelsdorf1995exact, sokal1998local, tiefelsdorf2002saddlepoint}.
Even in this literature, however, the efficiency of the \emph{estimator}
itself is usually not assessed, nor are alternative estimators for the
same estimand explored.

In this vein, \citet{wolf2025gisruk} recently showed that ordinary least
squares (OLS), the core estimating strategy for Moran-type statistics,
is a poor estimator choice for exploratory analysis. A single
sufficiently extreme observation (which we'll refer to as a
\emph{distributional outlier}) anywhere in the sample can drag the
Moran's \(I\) estimate to an arbitrarily distant value. This also
significantly affects local statistics. For instance, in the case of a
positive outlier (or increasing positive skew), an increasing fraction
of the data can become classified as ``low-low'' coldspot clusters. This
is concerning precisely because local Moran statistics are routinely
deployed to identify unusual observations and find anomalous areas of a
given map or image. Indeed, an estimator that can be \emph{misled} by a
single distributional outlier or skew should probably not be the default
tool for \emph{identifying} spatial outliers.

From a statistical perspective, the global aspect of this problem is
well-understood. The conventional global Moran statistic is a least
squares estimator, making it sensitive to distributional outliers.
Indeed, this implies it has a finite sample breakdown point of \(1/n\)
\citep{rousseeuw1987robust, maronna2006robust}. This applies even
without exploring the potential for \emph{configurational} effects
\citep{ma2000highly}, where an adversary can manipulate both what site
gets which corrupted value \emph{and} what other sites are connected to
the corrupted value. While the finite sample breakdown point of OLS is
commonly-understood, its relevance and application for local (and
global) indicators of spatial association have only recently begun to
recieve attention \citep{nardelli2025evaluating}. While geographers are
generally aware of the impact of spatial data quality issues such as the
perennial issue of overshoots/undershoots \citep{rey2026geoplanar} and
Null Island's continued popularity \citep{juhasz2022think}, the near
automatic use of paired Moran scatterplots and maps in exploratory
spatial data analysis is seriously challenged by this sensitivity.

Indeed, a more robust drop-in replacement estimator (instead of an
alternative estimand, \citet{li2007moran}) would be transformative,
given how common local Moran statistics are and their persistence in the
day to day practice of exploratory spatial data analysis. Like the
growth of permutation-based inference methods as drop-in replacements
for classical estimators
\citep{simon1999resampling, good2011practitioners, efron2016computer, simon2026resampling},
a drop-in robust replacement for the global and local Moran statistic
could see widespread use. Inference on these estimators is already
largely permutation-based \citep{sokal1993testing, sauer2021importance},
so robust and valid inference on such a drop-in robust estimator would
be immediately achieved.

Therefore, this paper studies and extends recent innovations in robust
spatial autocorrelation estimators over the last few years. We outline
\citet{arbia2025robust}'s Gnanadesikan-Kettenring and local ``plug-in''
robust estimators, \citet{wolf2025gisruk}'s Theil-Sen estimator, and
fully specify \citet{wolf2025gisruk}'s sketch of the \emph{trimmed least
squares} (TLS) Moran estimator (from \citet{rousseeuw2006computing}) for
the first time. This estimator iteratively censors distributional
outliers from consideration when representing spatial context and when
estimating spatial association. We propose a novel search procedure to
iteratively refine the trimmed fraction, in order to maximize the
efficiency of the statistic and improve its applicability for local
analysis, and outline a fast permutation-based inference technique. We
also clarify an important aspect of permutation inference with Theil-Sen
estimation of Local Moran statistics. Given these new plug-in,
Theil-Sen, and TLS indicators of local and global spatial association,
we run simulations across outlier- and skew-based spatial processes to
estimate the size, power, and performance of robust autocorrelation
statistics. We also define the paired Moran scatterplots for each of
these robust visualizations, and show a cartographic strategy for
visualizing TLS local statistics. Ultimately, we suggest that the
Theil-Sen estimator should replace the OLS Moran estimator in smaller
data, while the plug-in robust estimators can be used in larger data
where efficiency is less of a concern.

\section{Framing the Moran-Form
Regression}\label{framing-the-moran-form-regression}

To understand local and global measures of spatial association, it helps
first to begin with the Moran-form regression. In this setup, define
\(z\), a row-centered variate, and \(\mathbf{W}\), the \(N \times N\)
spatial weights matrix \citep{anselin1996moran} which records the
spatial linkages between observations. This might represent adjacency
relations between polygons within a lattice, points that neighbor one
another on a Delaunay or KNN graph, adjacent pixels in an image, or
connected points of interest in a city road network. \(\mathbf{W}\) is
also usually row-standardized, such that each row sums to one. Then, the
typical Moran-form regression is a procedure to estimate Morans' \(I\)
using a regression:
\begin{equation}\protect\phantomsection\label{eq-moran-reg}{
\mathbf{W}z \;=\; \alpha + I\,z + e
}\end{equation} The global \(I\) is the OLS slope for this regression:
\begin{equation}\protect\phantomsection\label{eq-moran-ols}{
\hat{I} = (z'z)^{-1}\,z'\mathbf{W}z \;=\; \frac{1}{n}\sum_i \sum_j z_i\, w_{ij}\, z_j
}\end{equation} This estimator is a \emph{global indicator of spatial
association}, which describes the covariation between observations and
the average of observations nearby. When positive, it indicates that
nearby observations are more similar to one another than if values were
distributed randomly across the map. From here, so-called \emph{local
indicators of spatial association} (LISA) \citep{anselin1995local} are
developed to describe the relationship between observations and their
neighbors at each site in the data. For the Moran statistic, local
statistics are obtained by switching the inner product to an elementwise
product: \begin{equation}\protect\phantomsection\label{eq-moran-local}{
  \hat{I}_i \;\propto\; z_i \cdot (\mathbf{W}z)_i
}\end{equation}

When considering the robustness of these estimators, we know that the
finite sample breakdown point for the estimator in
Equation~\ref{eq-moran-ols} is \(1/n\) since it is an OLS estimator.
However, because the Moran-form regression involves \(z\) on \emph{both
sides} of the regression, there are additional effects. To make the
consequences concrete: consider a single distributional outlier \(z_k\)
with \(|z_k| \gg \max_{j \neq k} |z_j|\). The outlier contributes both
directly (as \(z_k^2\) in the denominator and as \(z_k (\mathbf{W}z)_k\)
in the numerator) and indirectly (as \(w_{ik} z_k\) in the spatial lag
of every neighbour \(i \in \mathcal{N}(k)\)). As \(|z_k|\) grows, both
the global slope and the local statistics at \(k\)'s neighbours can be
driven to arbitrarily large or small values.In practice, the slope of
the Moran scatterplot is usually \emph{shallower} than it would be
otherwise if the outliers were properly identified and corrected. The
same applies for skewed distributions.

One formalization of this problem might involve the \emph{local
influence function} examined by \citet{arbia2026impact}. This
fundamental tool in robust statistics summarises the impact that any one
observation has on a statistic. \citet{arbia2026impact} correctly note
that ``the influence function of the Moran coefficient depends on the
contaminated value ... {[}and{]} also on the average of the values
observed in the neighborhood of the contaminated location.'' (p.~5).
While this recognizes the fact that outliers contaminate the process
both directly as sites and indirectly as lags, this focuses only on how
changes at a specific site affect global estimates, not necessarily how
the \emph{local statistic} itself enters into the global summary.

\section{Rethinking Robustness for Local
Statistics}\label{rethinking-robustness-for-local-statistics}

The \emph{local influence function} is useful, especially for spatial
econometric models where global estimates are the chief concern.
However, for local statistics, it is important to be clear that the
impact of \emph{distributional outliers}---observations whose values of
\(z\) are unusual no matter where in the map they sit---may differ from
that of \emph{spatial outliers}---observations whose values of \(z\) are
unusual only relative to their spatial surroundings. Distributional
outliers contaminate any estimator that uses \(z\); spatial outliers are
precisely the substantive signal that the LISA statistic is supposed to
detect. The local influence function does not distinguish the two.
Further, it does not \emph{as such} provide a strategy for creating
robust local indicators of spatial association that can keep a grip on
this distinction.

In light of this, a useful way to think about this is to decompose
robust LISA estimation into two coupled sub-problems that separate how
distributional outliers impact estimation:

\begin{description}
\item[(I) Local Association.]
How do we (robustly) summarise the site and its surroundings in the
presence of a distributional outlier?
\item[(II) Global Estimation.]
How do we (robustly) estimate the relationship between site values and
this summary in the presence of a distributional outlier?
\end{description}

\begin{figure}

\centering{

\pandocbounded{\includegraphics[keepaspectratio]{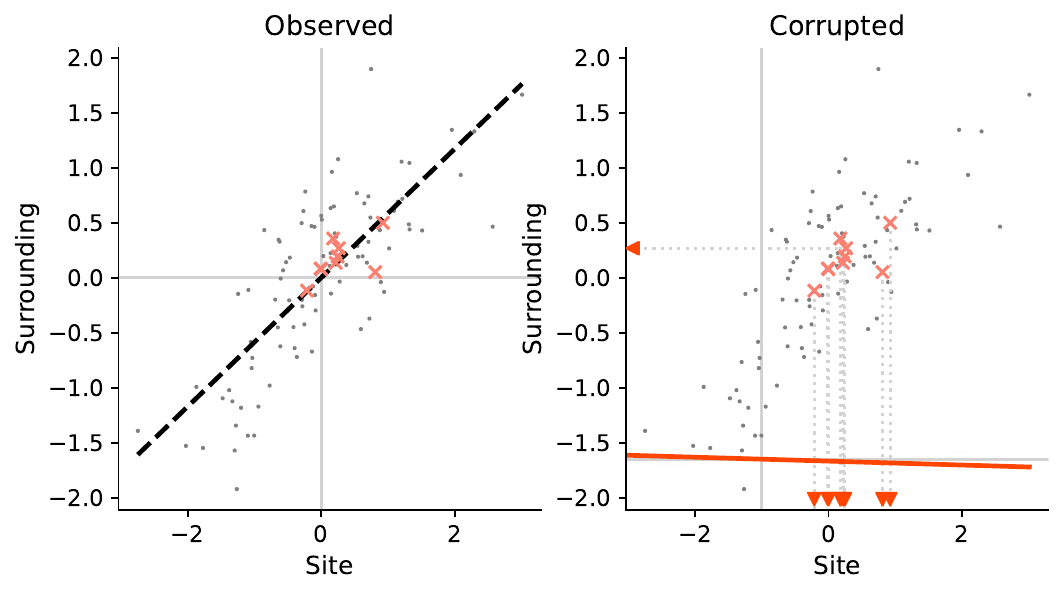}}

}

\caption{\label{fig-corruptplot}Moran Scatterplot of a
spatially-patterned random variable (left) and a corruption of the
pattern (right). Corruption of a single observation entails moving one
observation left or right. This, in turn, pushes other observations up
and down. Depending on the connectivity structure, manipulations of a
single observation adjust both the slope of the regression line (which
is the classical Moran's \(I\) estimate) as well as all local statistics
that are connected to the corrupted observation. Here, the corruption
entails moving an observation left by 100, which also pulls many other
observations downwards by adjusting their lag. For typical spatial
weights matrices, left-down and right-up are the only possible
directions for the corrupted observation and its neighbors' points.}

\end{figure}

In the classical Moran setup, (I) involves the row-standardised mean
spatial lag, \(\mathbf{W}z\), and (II) involves OLS. Both are
non-robust, in the sense that the lag of \(z\) is sensitive to
distributional outliers nearby, and OLS is sensitive to extreme sites
and extreme lags. Indeed, the two problems compound: an extreme \(z_j\)
enters its neighbours' lag (corrupting the local association), and it
also enters the regression directly through \(z_j\) itself (corrupting
the global estimation). Hence, a distributional outlier corrupts nearby
observations' vertical positions in the Moran Scatterplot in addition to
being an outlier itself on the horizontal axis. However, a spatial
outlier itself does not bias global estimation. This is apparent in
Figure~\ref{fig-corruptplot}, where spatial outliers (in the upper right
or bottom left quadrants) exist, but only corrupted data creates bias in
the right pane. Hence, any local statistic may be contaminated by
placing a corrupted within its neighbor set. Further, it is not
necessary for the ``corrupted'' observations to themselves start as
spatial outliers. Finally, any global estimate can also be contaminated
from a single site in typical network topologies.

\subsection{Solving Local Association and Global Estimation
separately}\label{solving-local-association-and-global-estimation-separately}

\citet{arbia2025robust} and \citet{nardelli2025evaluating} provide a
robust measure of spatial association using weighted medians for lag
calculations and median absolute deviations (MAD) for scale estimators
within a generalised correlation coefficient
\citep{gnanadesikan1972robust}. This ``plug-in'' strategy offers a
robust Moran estimator by replacing non-robust estimators of scale,
location, or covariance with robust estimators of the same estimand,
solving the local association and global estimation using separate
strategies.

\subsubsection{Problem (I): Robust Local
Association}\label{problem-i-robust-local-association}

The local association question concerns how outliers/skew corrupt
estimates of the ``typical value'' of surrounding sites; the mean is
sensitive to outliers/skew, so spatial lags get corrupted before any
association is ever attempted. To start, instead of standard
normalization, \citet{nardelli2025evaluating} use a robust normalization
with the median and MAD, \(z = (y-\text{median}(y))\text{MAD}(y)\) where
\(\text{MAD}(y) = \text{median}(|y - \text{median}(y)|)\). Then, a
plug-in strategy replaces the row-standardised mean of neighbours' \(z\)
with a more robust local summary---Hampel-style M-estimate, a trimmed
mean, or a weighted median over the neighbourhood. We'll denote this as:
\[
  \mathbf{W}^* z \;=\; \text{robust summary of } \{z_j : j \in \mathcal{N}(i)\}
\] \citet{nardelli2025evaluating} consider the (weighted) median for
their robust summary. That is, \(\mathbf{W}^*z\) calculates the
(weighted) median of \(z_i\) for each site \(i\) as the local
association. This addresses one direction of contamination---the
direction in which a neighbour's outlier value \(z_k\) pulls the
vertical axis of the Moran scatterplot for sites
\(i \in \mathcal{N}(k)\). After the fix, the robust summary at \(i\) is
unaffected by \(z_k\) unless \(k\) comprises a majority of \(i\)'s
neighbours, which is rare for typical contiguity graphs. Then, a plug-in
robust measure of local association is available as:
\begin{equation}\protect\phantomsection\label{eq-moran-plugin-local}{ 
\hat{I}_i^{\text{plug-in}} = z_i\circ[\mathbf{W}^*z]_i
}\end{equation} We call this the plug-in estimator for local
association.

\subsubsection{Problem (II): Robust Global Autocorrelation from Local
Statistics}\label{problem-ii-robust-global-autocorrelation-from-local-statistics}

From this point, one could create a global measure of spatial
autocorrelation using OLS with this robust lag:
\[\frac{\sum_i z_i \, (\mathbf{W}^* z)_i}{\sum_i z_i^2}\] Despite the
robust local association, this is not a robust global indicator. A site
whose own \(z_i\) is a distributional outlier still contributes
\(z_i^2\) to the denominator and \(z_i (\mathbf{W}^* z)_i\) to the
numerator. As \(|z_i|\) grows, the slope is pulled towards a line that
passes through \((z_i, (\mathbf{W}^* z)_i)\) regardless of the structure
of the rest of the data. Hence, the breakdown point along this second
axis remains \(1/n\).

An alternative might also be to construct the median-of-local site
estimators:
\begin{equation}\protect\phantomsection\label{eq-moran-median-of-locals}{
\hat{I}^{MOM} = \text{median}(z_i \circ [\mathbf{W}^*z]_i)
}\end{equation} While this estimator is straightforward to implement and
is clearly related to the local estimator, we focus instead on
\citet{nardelli2025evaluating}'s preferred estimator based on
\citet{gnanadesikan1972robust}'s nonparametric correlation. Defining
\(a^{-1}=MAD(z)\) and \(b^{-1} = MAD(\mathbf{W}^*z)\), we use the
polarization identity to convert the inner product between \(z\) and
\(\mathbf{W}z\) into a quotient relating the deviation of their sums and
differences: \begin{equation}\protect\phantomsection\label{eq-moran-gk}{
\hat{I}^{GK} = \frac{\text{MAD}(az + b\mathbf{W}^*z) - \text{MAD}(az - b\mathbf{W}^*z)}{\text{MAD}(az + b\mathbf{W}^*z) + \text{MAD}(az - b\mathbf{W}^*z)}
}\end{equation} This is a robust plug-in estimator for the global
correlation between \(\mathbf{z}\) and \(\mathbf{W}^*\mathbf{z}\) that
is functionally separate from the local robust estimate in
Equation~\ref{eq-moran-plugin-local}. Regardless, they can be used
together to define a robust Moran Scatterplot: \(I_{GK}\) estimates the
slope for a plot of \(\mathbf{W}^*z\) onto robustly-standardized \(z\)
through the origin, with \(I_i\) allowing for local inference on
observations in quadrants defined according to the medians of \(z\) and
\(\mathbf{W}^*z\).

The breakdown point of these estimators is somewhat complex since it
depends on the topology of \(\mathbf{W}\). Typical \(\mathbf{W}\)
matrices used spatial analysis are sparsely-connected, non-negative,
row-standardized, have minimum degree larger than 1, and are not
star-like. Under these conditions, these estimators have a 50\% finite
sample breakdown point: a single corrupted site is isolated from nearby
sites through the robust spatial lag, and the global estimator is
isolated from the corrupted site by the median absolute deviation.
Regardless of the breakdown point, though, we expect these statistics to
be somewhat underpowered, especially in skewed distributions
\citep{rousseeuw1993alternatives}.\footnote{This is where
  \citet{arbia2025robust}'s comparison with Box-Cox de-skewing
  transforms becomes quite interesting. This is deserving of further
  study as a central concept, and we defer it here.} Thus, we explore
other estimators for local and global spatial association that have
higher power.

\subsection{Theil-Sen Regression Addresses Local and Global
Robustness}\label{theil-sen-regression-addresses-local-and-global-robustness}

\begin{figure}

\centering{

\pandocbounded{\includegraphics[keepaspectratio]{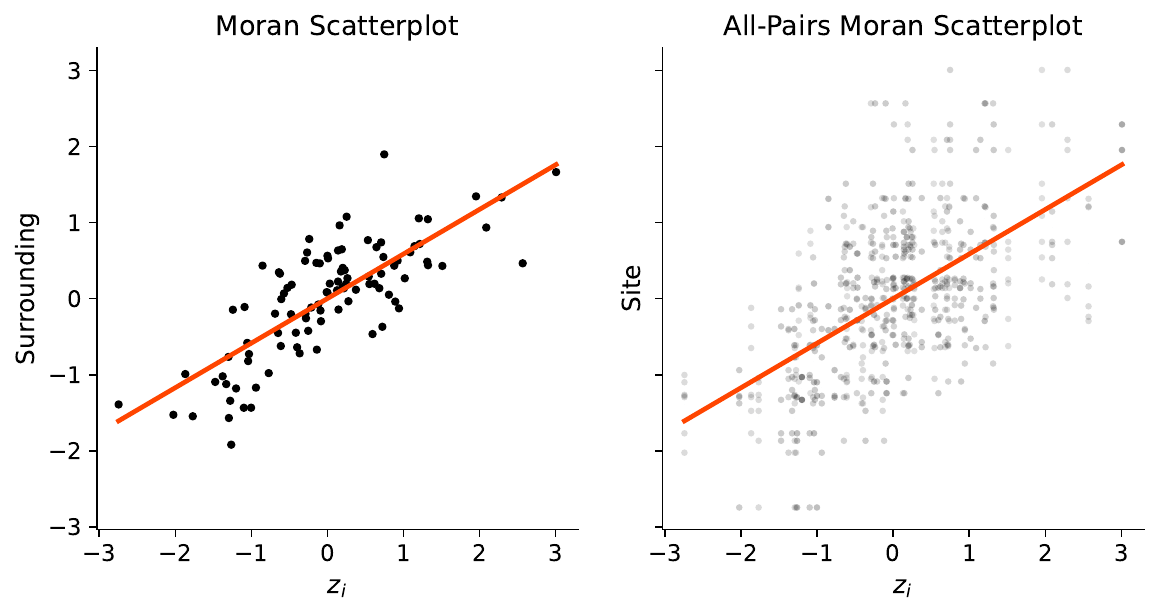}}

}

\caption{\label{fig-allpairs}Comparison between the classic Moran
Scatterplot (left) and the implied all-pairs weighted linear regression
(right), where pairs of points \(z_i,z_j\) are made darker proportional
to their \(w_{ij}\) value.}

\end{figure}

To improve power, it is useful to explore robustness strategies informed
by the wider (geo)statistical literature. For instance, estimators of
\citet{dowd1984variogram} and \citet{genton1998highly} are fairly
efficient. They consider quantiles of pairwise differences, which
improves their general performance. The nonparametric estimator of
\citet{bjornstad2001nonparametric} also examines fitting general
functions of pairwise relationships in spatial data. In this vein,
\citet{wolf2025gisruk} argue that instead of plug-in estimators, a
robust regression-based estimator can solve (I) and (II) simultaneously.
However, this requires us to change the framing of the problem from a
robust regression on robust local associations to solving a single
robust weighted regression across all pairs of observations, as
visualized in Figure~\ref{fig-allpairs}. The local part refers to the
relation between a robust regression line and the vertical stripe of
points at each \(z_i\) value; the global part refers to the slope of the
robust regression line itself. This also clarifies how a contaminated
observation \(z_k\) shows up as both the horizontal coordinate (in pairs
\((z_k, z_j)\)) \emph{and} the vertical coordinate (in pairs
\((z_i, z_k)\)) of many points in the scatter. To estimate this
regression line in a robust fashion, \citet{wolf2025gisruk} define a
Theil-Sen iterated medians estimator for local and global Moran
statistics: the local association is robustly estimated using the
weighted median of slopes constructed from pairs of points located at
\(z_i\), while the global estimation is addressed using weighted median
of local medians.

To provide more detail on this estimation strategy, let's define \(Z^i\)
and \(Z^j\) as two matched vectors of \(z\), where the \(k\)th entry of
each describes one pair \((i, j)\). Let \(\tilde{\mathbf{W}}\) be the
diagonal matrix of pairwise weights, with
\(\tilde{\mathbf{W}}_{kk} = w_{ij}\). The weighted-least-squares
estimator \begin{equation}\protect\phantomsection\label{eq-allpairs}{
  \hat{I}^{\text{WLS}} = (Z^{i\prime}\tilde{\mathbf{W}}Z^i)^{-1}\, Z^{i\prime}\tilde{\mathbf{W}}Z^j
}\end{equation} returns the same value as Equation~\ref{eq-moran-ols}
for typical \(\mathbf{W}\). This is because the two are unbiased
estimators of the same estimand: the relationship between \(z_i\) and
the values of \(z_j\) in the rest of the map, weighted by spatial
proximity. The classical estimator first averages \(z_j\) over
\(j \in \mathcal{N}(i)\) and then runs OLS on that average; the
all-pairs estimator considers all pairs \((z_i, z_j)\) at once,
weighting each pair by \(w_{ij}\).

As a robust alternative to WLS, the Theil-Sen-Moran estimator (with
Siegel-style iterated means) can be used. The Theil-Sen Local Moran
estimator at each site is the median of site-specific slopes (shown in
Equation~\ref{eq-ts-local}) and the global estimator is the median of
site medians (shown in Equation~\ref{eq-ts-global}). To obtain these
estimates, we calculate all of the slopes from \((z_i,z_j)\) to any
other point \((z_k,z_l)\) for each \(i\) such that neither
\(j \neq k,l\) and \(i \neq k,l\):
\[s_{ijkl} = \frac{z_l-z_j}{z_k-z_i}\] The weight of this slope is
\(w_{ijkl} = w_{ij}w_{kl}\). We then calculate the local Theil-Sen slope
at \(i\) as the weighted median of these slopes over \(jkl\):
\begin{equation}\protect\phantomsection\label{eq-ts-local}{
\hat{I}_i^{TS} = \text{median}(s_{ijkl},\,w_{ijkl}) 
}\end{equation} From here, we calculate site weights as
\(w_i = \sum_{jkl}w_{ij}w_{kl}\), and estimate the global slope as the
weighted median of site-level slopes:
\begin{equation}\protect\phantomsection\label{eq-ts-global}{
\hat{I}^{TS} = \text{median}(\hat{I}_i,\,w_i)
}\end{equation}

As noted above, this addresses the local association and global
estimation problems simultaneously during estimation with the iterated
weighted medians.\footnote{One might instead argue that an explicit
  solution to the local association question is \emph{never
  constructed}, since the lag is never computed. However, our definition
  of the local association problem is the \emph{relation between site
  and surrounding}, not \emph{the surrounding} value itself. This is
  provided by the slope.} This achieves a breakdown point of 50\% over
typical graphs, avoids introducing any hyperparameter, does not involve
the MAD, and is quite efficient. Like the plug-in estimator, a robust
Moran scatterplot can be defined using medians, or using the robust
Theil-Sen intercept estimate.

This estimator has a few drawbacks, however. Notably, the Theil-Sen
estimator scales poorly with both sample size and graph density. While
the calculation can be easily vectorised and parallelised (and is thus
made amenable to common high performance computing patterns) the
fundamental scaling limitations are made immediately apparent when
conducting permutation inference on the local slopes, which requires a
massive number of iterations of the estimator. As \(N\) becomes truly
large (\(N>100,000\)) or graphs become dense (as when estimating
autocorrelation over non-compact kernels), this simply becomes
infeasible. In addition, the local median slope is also not a
cross-product, so it differs slightly from the interpretation of the
classical Local Moran's \(I\). That said, it is not advised to compare
the raw \(I_i\) statistics directly, since they depend so strongly on
the unique topological structure of the graph on which they are
estimated; it is more common to convert them into \(z\)-scores, or to
compare local statistics only in terms of quadrant classifications and
permutation \(p\)-values. The Theil-Sen Local Moran admits both.

\subsection{Trimmed Least Squares for (Local) Moran
Estimation}\label{trimmed-least-squares-for-local-moran-estimation}

\citet{wolf2025gisruk} acknolwedged these drawbacks, and speculated that
the TLS estimator \citep{rousseeuw2006computing} may improve upon the
Theil-Sen iterated medians estimator. Therefore, we investigate whether
the TLS Moran statistics can provide fast robust local and global
indicators of spatial association statistics with similar semantics to
the classic Moran statistics. To formalize this approach, the usual TLS
problem is defined as an optimization problem to finding
\(\alpha,\beta\) that minimize the sum of squared residuals among
\(N(1-q)\) retained observations, where \(q\) is some fraction of the
data to be trimmed:
\begin{equation}\protect\phantomsection\label{eq-trimmed-ls}{
\begin{aligned}
\underset{\alpha,\beta}{\text{minimize}}& \sum_i \biggr(h_i\big(y_i - \alpha - X_i\beta\big)\biggr)^2\\
\text{subject to:}&\\
&\sum_i h_i \geq N(1-q) \\
&h_i \in {0,1}
\end{aligned}
}\end{equation} for \(N \times 1\) response vector \(y\), \(N \times p\)
covariate vector \(X\), \(h_i\) binary decision variable equal to \(1\)
when \(i\) is not trimmed, and \(q \in [0, 1/2)\) as the trim fraction.
This is a quadratic integer program with a single linear constraint. It
is tractable in small data, and heuristic solutions are quite effective
in larger data \citep{rousseeuw2006computing}.

To extend this to a Moran-form regression, we must re-define the \(y\)
component of the problem as a function of \(z\). To help, let us define
a re-normalizing lag operator, which adjusts the lag calculation to
account for trimmed observations:
\begin{equation}\protect\phantomsection\label{eq-trimmed-lag}{
R(z_i) = \sum_j \frac{w_{ij}h_jz_j}{\sum_j w_{ij}h_j}
}\end{equation} This is the spatial lag at site \(i\) only among
survivors. With this, the TLS Moran optimization problem is:
\begin{equation}\protect\phantomsection\label{eq-tls-objective}{
\begin{aligned}
\text{minimize}& \sum_i \biggl(h_i\bigl(R(z_i) - \alpha - z_i \beta\bigr)\biggr)^2\\
\text{subject to:}&\\
&\sum_i h_i \geq Nq \\
&h_i \in {0,1}
\end{aligned}
}\end{equation} The identified \(\hat{\alpha}\) for this approach
represents the robust intercept estimate, and the
\(\hat{I}_i^{TLS} = \hat{\beta}\) represents our TLS global indicator of
spatial association. Local estimates can then be obtained from
\(\hat{I}_i^{TLS} = z_i\circ R(z_i)\). The finite sample breakdown point
for the TLS global indicator is governed by \(q\): if \(q=.5\), then the
breakdown point is 50\%. The TLS approach is also relatively efficient
if there are few outliers, but loses the efficiency as the trimming
fraction increases. As a unique strength, the TLS method explicitly
identifies a subset of distributional outliers as a side-effect of the
estimation. This extends the typical Local Moran quadrant-based
classification to include values that would be considered distributional
outliers regardless of where in the map they are located. Alternatively,
these distributional outliers can be re-contextualized by evaluating the
counterfactual: \emph{if these were not trimmed, what classification
would they be?} This is discussed at length in
Section~\ref{sec-counterfactual-tls}. Finding a good trimming fraction
(\(q\)) and ultimately estimating the model are also challenging, and
our heuristic approaches for doing so are discussed in
Section~\ref{sec-tls-solving} and Section~\ref{sec-trimming-fraction}.

\subsubsection{Conditional Permutation Inference for Robust Local
Inference}\label{sec-counterfactual-tls}

The conditional permutation procedure of \citet{ord1995local} for local
statistical inference is a commonly-employed method to conduct inference
on local indicators of spatial association of any form
\citep{sauer2021importance}. This strategy constructs a pseudo
\(p\)-value at each site \(i\) by holding \(z_i\) fixed, randomly
exchanging the values of neighbors into the other \(n-1\) sites, and
recomputing the local indicator at \(i\). Repeating this \(R\) times
produces a reference distribution at each site, against which the
observed local statistic is compared. This approach can be used as is
for the plug-in estimators. For the local Theil-Sen indicator, it is
important to make sure that we must hold \(i\) fixed and exchange
\emph{only} \(i\)'s neighbors' values with the rest of the map. Holding
\(i\) fixed and shuffling the entire rest of the map unduly (and
erroneously) breaks structure that the Theil-Sen local statistic is
sensitive to, through the all-pairs median of medians. The classic Moran
statistic is insensitive to quite a few different permutation
configurations \citep{sokal1998local}, but the local Theil-Sen is not a
Mantel-type crossproduct statistic
\citep{mantel1967detection, hubert1981generalized}. Therefore, it
requires shuffling only the neighbors into the rest of the
map---shuffling all values except \(i\) for inference at \(i\) results
in spurious detections.

The simplest way to adapt the permutation inference strategy for
TLS-derived indicators is to repeat the TLS estimation during each
permutation, whether conditional (site \(i\) fixed while neighbors are
permuted, local null) or total (all sites permuted, global null).
However, this presents a few complications. Since trimming decisions
take into account the spatial pattern through \(\mathbf{W}z\), the
trimmed set may be different in each conditional permutation, as well as
identified \(q\) (if not held fixed). For the global statistic,
calculation of pseudo-\(p\) values is straightforward using
\((M+1)/(R+1)\), where \(M\) is the count of those equal to or more
extreme than the observed statistic. For local statistic, we can either
simulate \(R\) conditional permutations and use only surviving
replications to obtain \((M+1)/(R'+1)\), where \(R'\) is the number of
valid (non-trimmed) permutations, or simulate conditional permutations
at each site until \(R'=R\).

This repeat-survivor approach is simple theoretically, but is
exceptionally slow. If we fix the number of replicates per site, then we
run a potentially unbounded number of expensive TLS estimations. This
will make TLS impractical in all but the smallest of applications.
Second, what a ``trim'' represents is not clear in this framework.
Arguably, a site that is trimmed in many null permutations is a site
whose local statistic is hard to characterise under the null. The size
of the reference distribution under the null (or the number of
permutations required to obtain exactly \(R\) null replicates) is itself
informative about the local indicator. Indeed, the repeat-survivor
permutation \(p\)-value implicitly contains two hypotheses:
\emph{conditional upon} \(i\) not being an outlier, does \(i\) have
null-defying local structure? This null is of dubious relevance to sites
that were in fact trimmed.

To avoid this massive amount of computation for uncertain benefit, we
suggest that a counterfactual null for trimmed sites is more appealing.
To achieve this, we treat the trimmed set as fixed and shuffle only
survivor values for inference. Global and local inference on survivors
looks similar to the shuffling part of the repeat-survivor approach, but
local inference on trimmed sites is quite different. For trimmed sites,
we calculate local statistics under the null using survivor lags
\([\mathbf{W}^*z]_i\). The resulting local permutation distribution
allows us to infer the local structure around the site if it had not
been trimmed. This allows us to inspect the local association around
trimmed observations, while repeat-survivor \(p\)-values are only
reliable if \(R'\) is large. Second, the counterfactual procedure will
be \emph{much faster} than repeating the TLS algorithm for every
\(N \times P\) local permutation. However, only the repeat-survivor
approach represents the uncertainty about which observations enter the
trimmed set. Further, it is not immediately clear whether the two
techniques are likely to agree, especially in cases where \(q\) is
identified from the data. Therefore, we investigate whether these
results lead to substantively different interpretations in our
simulation design.

\subsubsection{Solving the trimming problem with C-step
estimation}\label{sec-tls-solving}

The structure of Equation~\ref{eq-tls-objective} is distinct from the
classical least-trimmed-squares problem in Equation~\ref{eq-trimmed-ls}.
In the classical TLS problem, \(y\) and \(X\) are fixed during solving;
changing which observations survive trimming does not alter their \(y\)
or \(X\) values. In contrast for TLS Moran, response \(R(z_i)\) is
itself a function of \(h_j\); dropping observation \(j\) changes
\(R(z_i)\) for every \(i\) that has \(j\) as a neighbour. This, in turn,
changes the residuals of those rows. Hence,
Equation~\ref{eq-tls-objective} is a unique combinatorial problem in
which the objective function itself changes as the mask changes. Solving
it exactly is intractable except at very small \(n\), and a heuristic is
needed.

We adapt the C-step algorithm of \citeauthor{rousseeuw1999fast}
\citetext{\citeyear{rousseeuw1999fast}; \citealp{rousseeuw2006computing}},
since it is a fast heuristic for the standard LTS problem. Given a
current mask \(H^{(t)}\), the standard C-step fits OLS on rows in
\(H^{(t)}\), computes squared residuals for all rows, and selects the
\(h\) rows with smallest squared residuals to form \(H^{(t+1)}\). The
classical result is that the trimmed loss is monotonically
non-increasing across iterations, and the algorithm converges in
finitely many steps to a local minimum. Multiple random restarts
approximate the global optimum. The spatial variant modifies steps
(1)--(3) to account for the dependence of the response on \(H\):

\begin{enumerate}
\def\labelenumi{\arabic{enumi}.}
\tightlist
\item
  Compute the trimmed spatial lag \(R(z)\) from
  Equation~Equation~\ref{eq-trimmed-lag}
\item
  Fit OLS of \(R(z)\) on \(z\), restricted to rows \(i \in H^{(t)}\)
  with at least one surviving neighbor.
\item
  Compute squared residuals for all rows \(i\) with at least one
  surviving neighbor;
\item
  Select the \(h\) rows with smallest squared residuals to form
  \(H^{(t+1)}\).
\end{enumerate}

The classical argument for monotonically-decreasing loss does not carry
over to the Moran variant. When an observation enters or exits the
trimmed set between iterations, the response variable changes for its
neighbours, which in turn changes their residuals; a row that had a
small residual at iteration \(t\) may have a large residual at iteration
\(t+1\) even if the regression coefficients do not move much. So the
C-step in our setting is a heuristic without a global-convergence
guarantee. In practice, we observe convergence in a small number of
iterations---typically fewer than ten---for the problem sizes considered
here, and we use multiple random starts to mitigate the loss of
monotonicity.

\subsubsection{How many observations to
trim?}\label{sec-trimming-fraction}

One additional complication in using trimming in this context is that
\(q\) is a hyperparameter which must be set in advance. Especially in
large data, there is no harm in setting \(q=.5\). In this instance, the
counterfactual quadrant labels for trimmed observations are likely
useful. However, there may be significant topological effects on
\(\mathbf{W}\) from trimming so many observations---we might create many
isolates or swaths of the map where no observations are retained. This
might make it impossible to apply the counterfactual inference
procedure, or may seriously thin the degree distribution such that the
variance of local statistics becomes unacceptably large. Therefore, we
suggest a simple automatic procedure for identifying suitable \(q\).
Calculate estimates at the maximum trim (\(q=.5\) obtaining
\(\hat{\beta}_{.5}\)) and the minimum trim (\(q=0\) obtaining the OLS
\(\hat{\beta}_0\)). Then, conduct a forward search starting from \(q=0\)
for the smallest trimming fraction \(\hat{q}\), such that trimming
\(\hat{q}\)\% of the data obtains \(\hat{\beta}_q\) within at least
\(\hat{q}\)\% of \(\hat\beta_{.5}\):

\[\frac{|\hat{\beta_q} - \hat{\beta}_{.5}|}{\hat{\beta}_{.5}} \leq \hat{q}\]

For instance, if trimming 5\% of the data obtains \(\beta\) within 1\%
of \(\beta_{.5}\), we terminate. This represents a search cone anchored
at \(q=0\), providing increasing tolerance for deviation from
\(\beta=.5\) as we trim more observations. This is useful for us to find
a \emph{small} \(\hat{q}\) that gets close to the maximally-robust
\(q=.5\), but by no means provides a minimal \(q\) such that
distributional outliers are always identified.

\section{Simulation Design}\label{simulation-design}

To demonstrate these new techniques, we use a simulation experiment
using spatially-patterned random variables with controlled skew and
kurtosis. For the latter, we run experiments using a SARTRE
distribution, or \emph{SAR with \(t\)-distributed random effects}. This
process takes a typical SAR random variate and replaces the
normally-distributed error with a matrix-\(t\)-distributed error. Define
the filter matrix \(\mathbf{F}=(I - \rho \mathbf{W})\), then define the
SARTRE as: \begin{equation}\protect\phantomsection\label{eq-sartre}{
  \mathrm{SARTRE}(\nu, \rho) = F(\rho)^{-1}t_n(\nu)
}\end{equation} Here, \(\rho\) governs the spatial autocovariance in the
process and \(\nu\) defines the thinness of tails. As
\(\nu\rightarrow\infty\), the SARTRE process converges to a classical
SAR process. As \(\nu\) approaches \(1\), the tails of the distribution
get heavier. Alternatively, to test skew, we can define a
\(\mathrm{SARLN}(\sigma,\rho)\) distribution similarly:
\begin{equation}\protect\phantomsection\label{eq-sarln}{
  \mathrm{SARLN}(\sigma, \rho) = F(\rho)^{-1}\mathrm{Lognormal}_n(\sigma)
}\end{equation}

We use these to consider the separate impacts of skew and outlier
frequency on these robust local statistics, as well as the classic local
statistic. As \(\sigma\) increases in the SARLN, the right skew
increases. As \(\nu\) decreases in the SARTRE distribution, kurtosis
increases. In the degenerate case of \(\nu=1\), the distribution
collapses to a Cauchy distribution which has infinite variance.

We define a simulation grid for the SARTRE and SARLN processes with
\(\nu\in{1,2,5}\) and \(\sigma\in{.5,1.5,3}\). We also include a
standard SAR simulation at each \(\rho\)/\(n\) to represent
\(\sigma\rightarrow0\) and \(\nu\rightarrow\infty\). We set
\(n\in{100,1000,10000}\), and \(\rho \in {0, .25, .5, .75, .95}\). We
run 99 conditional permutations for the local statistics, and 99 global
permutations for the global statistic across two samples sizes,
\(n=100\) and \(n=1000\).

Overall, simulations at \(\rho=0\) allow us to assess statistical size:
how frequently does the global statistic identify a trend when one does
not exist? Simulations at \(\rho \neq 0\) allow us to assess statistical
power: how frequently does the statistic identify a trend when one does
exist? If the power decreases with \(\sigma\), the estimator is not
robust to skew. If the power decreases with \(\nu\), the estimator is
not robust to kurtosis/outliers. We also consider the bias of estimates
relative to the classic Moran's \(I\) in cases where
\(\sigma\rightarrow0\)/\(\nu\rightarrow\infty\). The efficiency of
estimators is measured in terms of their timing as a function of sample
at a given weights matrix sparsity level, and as a function of weights
matrix sparsity at a fixed sample size. To illustrate this simulation
process, we show a \(SARLN(\sigma=1,\rho=.5)\) example first. Then, we
discuss results across all realisations.

\subsection{Expected Outcomes from Simulations}\label{sec-expectations}

Given these processes, we expect the following to happen over
simulations from these distributions:

\begin{enumerate}
\def\labelenumi{\arabic{enumi}.}
\item
  at \(\sigma=0\)/\(\nu\rightarrow\infty\), global statistic estimates
  will differ. We anticipate the Theil-Sen to be closest to the
  classical Moran statistic across all simulation configurations.
\item
  at \(\rho=0\), statistics should have appropriate size: \(p\) values
  over replicates should be uniform for both local and global
  statistics, with \(.05\) rejecting the null at a \(95\)\% confidence
  level.
\item
  at \(\rho \neq 0\), global statistic estimates will differ in power.
  We expect the Theil-Sen to be the most powerful estimator across all
  \(\rho \neq 0\). We expect the TLS to be next-most efficient, followed
  by the plug-in for robust global estimators.
\item
  for increasing \(\sigma\) and decreasing \(\nu\), we expect the
  classical Moran to lose power and generate deflated global estimates.
\item
  as \(\sigma\) increases in the \(SARLN\), we expect more \(LL\) local
  statistics for the classical Moran statistic.
\item
  Theil-Sen local statistics will agree most closely with the plug-in
  robust local association measure, and anticipate that the local TLS
  local statistic is maximally similar to the classical local Moran
  statistic on average over realizations.
\item
  Theil-Sen will be the slowest to compute, followed by TLS, plug-in,
  and then classical Moran.
\item
  The TLS counterfactual inference procedure will yield similar
  \(p\)-values to the repeated-survivor's conditional \(p\)-value at
  substantially lower computational cost.
\item
  Detected \(q\) for the TLS statistic will be larger when skew is
  stronger (\(\sigma\) is large) and distributional tails are fatter
  (\(\nu\) is smaller).
\end{enumerate}

\section{Results}\label{results}

To investigate the presence of these models, it helps to first consider
a realisation from the simulation set. Then, we present the results from
all simulations. Finally, we present the results from a specific
application.

\subsection{Indicative Simulation
Outcome}\label{indicative-simulation-outcome}

\begin{figure}

\centering{

\pandocbounded{\includegraphics[keepaspectratio]{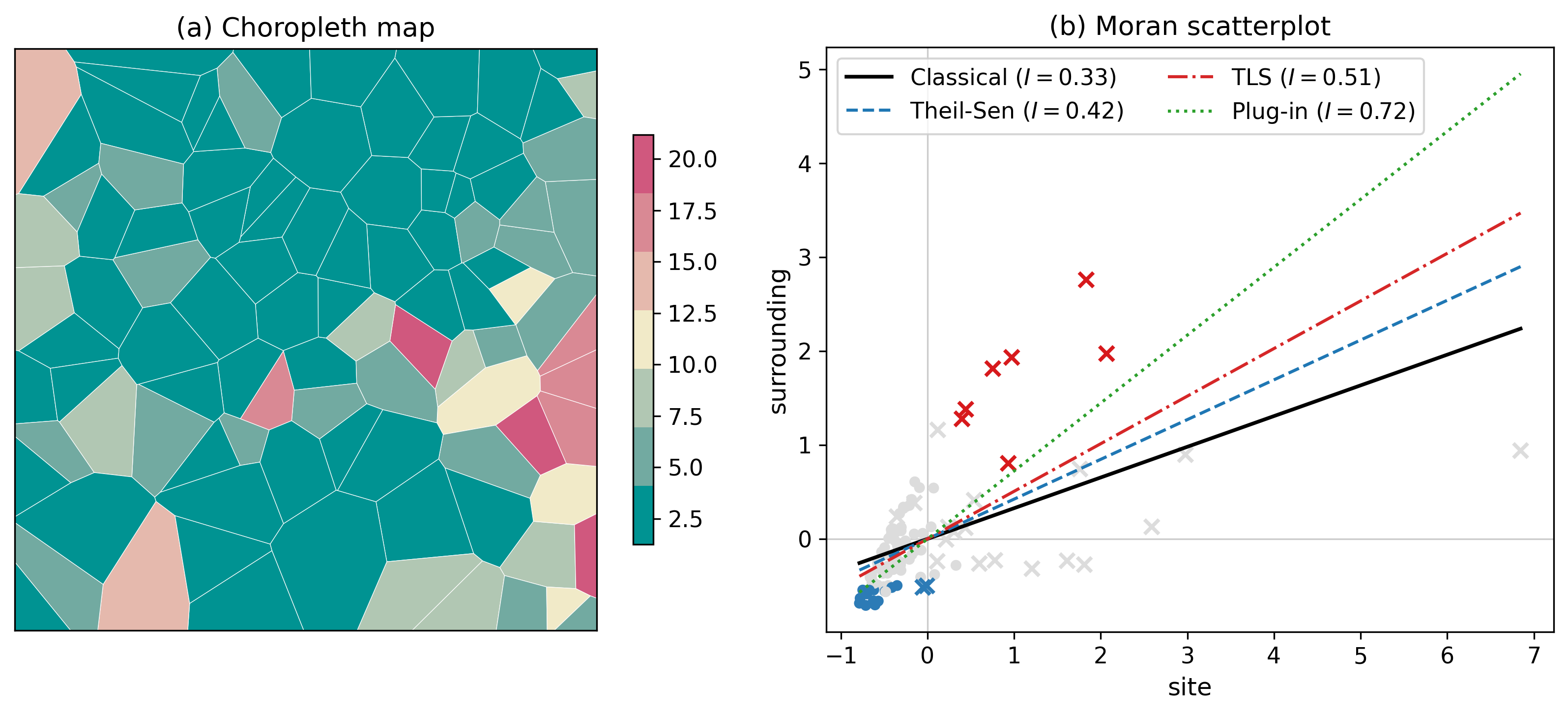}}

}

\caption{\label{fig-single-realisation}(a) One realisation of
\(SARLN(\sigma=1.5,\rho=0.5)\) at \(n=100\), drawn as choropleth map
over Voronoi cells. (b) The corresponding Moran scatterplot, with sites
coloured by their classical local-Moran cluster and the OLS slope
(global Moran's \(I\)) drawn through the cloud.}

\end{figure}

To build intuition before turning to the full simulation,
Figure~\ref{fig-single-realisation} shows a single realisation of the
\(SARLN(\sigma=1.5,\rho=0.5)\) process at \(n=100\). The map of values
(panel (a)) is dominated by a handful of large draws from the lognormal
tail; because the process is spatially autoregressive, these large
values bleed into their neighbours, producing the high-leverage points
in the upper-right of the Moran scatterplot (panel (b)). The classical
global Moran's \(I\) for this realisation is \(0.33\), which is the
lowest global estimate. The Theil-Sen estiamtor is closest to the
classical estimator \((.42)\), while the TLS is close to \(\rho\)
(\(.51\)) in value\footnote{We do not expect \(E[\hat{I}]=\rho\) for
  this process, so this is not necessarily a benefit.} and plug-in is
quite high (\(.72\)) relative to other estimates.

\begin{figure}

\centering{

\pandocbounded{\includegraphics[keepaspectratio]{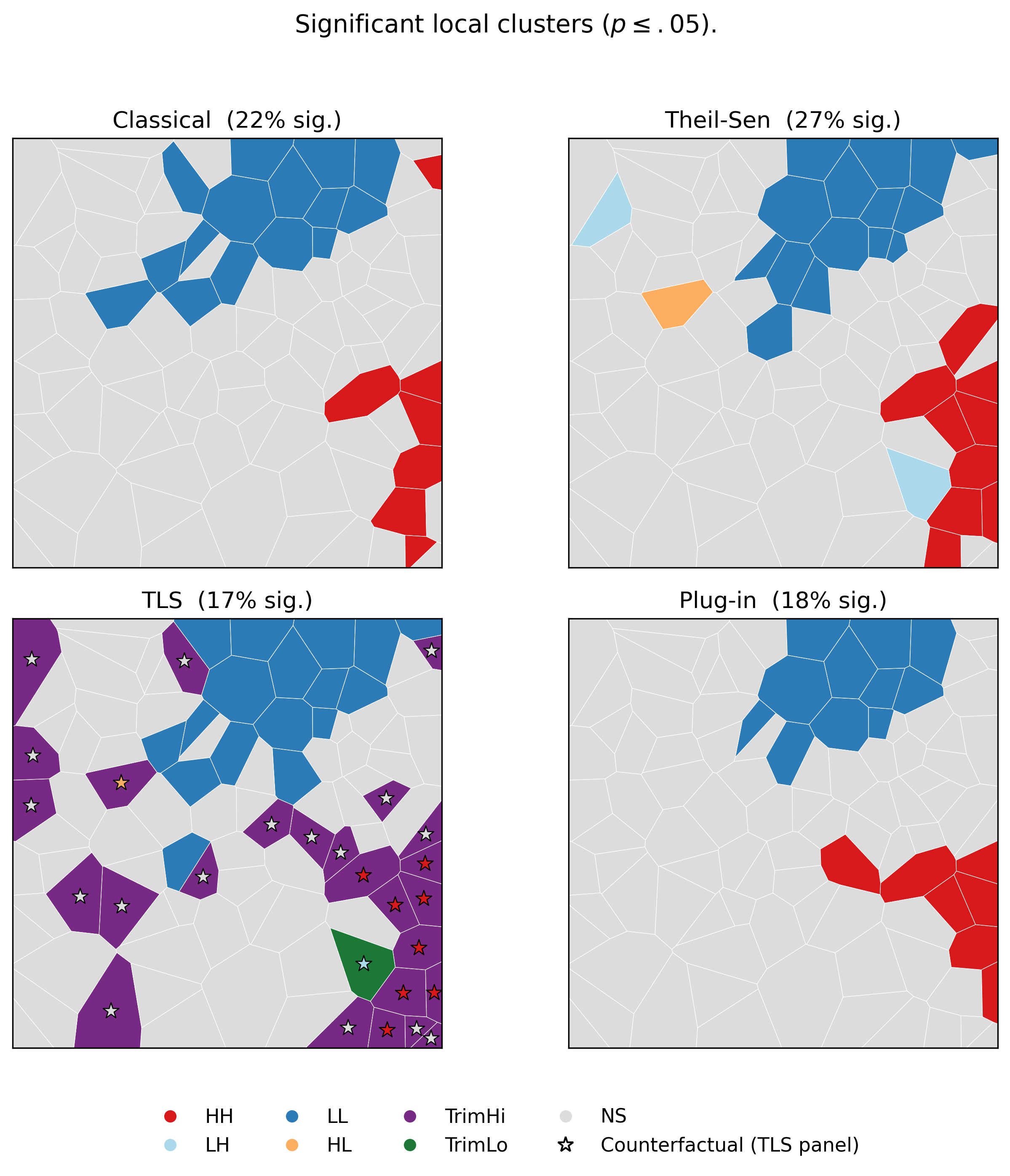}}

}

\caption{\label{fig-single-clusters}Significance-filtered local cluster
maps (\(p\leq0.05\)) for the realisation in
Figure~\ref{fig-single-realisation}, one Voronoi panel per estimator.
\texttt{TrimHi}/\texttt{TrimLo} mark sites the TLS removes as
high-leverage before fitting.}

\end{figure}

Figure~\ref{fig-single-clusters} maps the significant (\(p\leq0.05\))
local clusters for each estimator, and Table~\ref{tbl-crosstab-ts},
Table~\ref{tbl-crosstab-gk}, and Table~\ref{tbl-crosstab-tls}
cross-tabulate the classical cluster assignment against each robust
estimator. The Theil-Sen flags the most sites overall (27, against the
classical 22 and the TLS 17), while the plug-in is the most conservative
here (18). The classical statistic, Theil-Sen, and global plug-in each
report a high-high cluster (7, 9, and 7 sites, respectively) with
somewhat overlapping geometries. The TLS, however, reports no hotspot/HH
clustering. Instead, the TLS trims 27 of the most extreme values, only
one of which occurs in the left (lower) tail. All HH sites detected
across the classical, Theil-Sen, or plug-in local statistics are
trimmed, except for a single HH plugin site towards the center. All four
estimators agree closely on the low-low cluster in the upper-right of
the map, but differ at the edges. The Plug-In estimator recovers an LL
coldspot as a subset of the classical statistic, while the others do
not. The Theil-Sen is the only site that identifies spatial outliers,
but it does so in a way that largely comports with visual intuition in
the value map. For example, the high-low cluster is larger than all of
its neighbors. The TLS procedure trims this observation, but gives it a
counterfactual classification as a HL spatial outlier as well. The sole
lower-tail distributional outlier in the TLS case is identified as an LH
spatial outlier, as in the Theil-Sen case. Thus, there is a broad
agreement between the various statistics about cluster cores, but
cluster geometries differ on the margins. Further, the Theil-Sen
estimator seems more likely to identify spatial outliers, and some of
these comport with distributional outliers identified by the TLS
technique. The TLS technique, on its own, appears to trim too much,
although the counterfactual classification can be used to mitigate the
effect of this trimming.

{

\begin{longtable}[]{@{}lrrrrr@{}}

\caption{\label{tbl-crosstab-ts}Classical local-Moran cluster (rows)
against the Theil-Sen Moran assignment (columns), single
\(SARLN(\sigma=1.5,\rho=0.5)\) realisation at \(n=100\).}

\tabularnewline

\toprule\noalign{}
Classic & HH & LH & LL & HL & NS \\
\midrule\noalign{}
\endhead
\bottomrule\noalign{}
\endlastfoot
HH & 5 & 0 & 0 & 0 & 2 \\
LH & 0 & 0 & 0 & 0 & 0 \\
LL & 0 & 0 & 10 & 1 & 4 \\
HL & 0 & 0 & 0 & 0 & 0 \\
NS & 4 & 2 & 5 & 0 & 67 \\

\end{longtable}

}

{

\begin{longtable}[]{@{}lrrrrr@{}}

\caption{\label{tbl-crosstab-gk}Classical local-Moran cluster (rows)
against the plug-in (GK) Moran assignment (columns), single
\(SARLN(\sigma=1.5,\rho=0.5)\) realisation at \(n=100\).}

\tabularnewline

\toprule\noalign{}
Classic & HH & LH & LL & HL & NS \\
\midrule\noalign{}
\endhead
\bottomrule\noalign{}
\endlastfoot
HH & 4 & 0 & 0 & 0 & 3 \\
LH & 0 & 0 & 0 & 0 & 0 \\
LL & 0 & 0 & 11 & 0 & 4 \\
HL & 0 & 0 & 0 & 0 & 0 \\
NS & 3 & 0 & 0 & 0 & 75 \\

\end{longtable}

}

{

\begin{longtable}[]{@{}lrrrrrrr@{}}

\caption{\label{tbl-crosstab-tls}Classical local-Moran cluster (rows)
against the TLS Moran assignment (columns), single
\(SARLN(\sigma=1.5,\rho=0.5)\) realisation at \(n=100\).
\texttt{TrimHi}/\texttt{TrimLo} count sites the TLS removes as
high-leverage.}

\tabularnewline

\toprule\noalign{}
Classic & HH & LH & LL & HL & NS & TrimHi & TrimLo \\
\midrule\noalign{}
\endhead
\bottomrule\noalign{}
\endlastfoot
HH & 0 & 0 & 0 & 0 & 0 & 7 & 0 \\
LH & 0 & 0 & 0 & 0 & 0 & 0 & 0 \\
LL & 0 & 0 & 13 & 0 & 0 & 2 & 0 \\
HL & 0 & 0 & 0 & 0 & 0 & 0 & 0 \\
NS & 0 & 0 & 4 & 0 & 56 & 17 & 1 \\

\end{longtable}

}

The robust Moran scatterplot is visualized below in figure
Figure~\ref{fig-robust-scatterplot}. Both the plug-in and Theil-Sen
estimators use the same axis structure, while the TLS scatter uses the
trimmed mean/trimmed lag mean for axes. Clearly, the strong trimming by
the TLS estimator reduces the spread of observations vertically,
compressing the structure of the scatterplot. Regardless, we can see
that points are at the same locations for the plug-in and Theil-Sen
robust Moran scatterplots, but the line (and significance decisions)
will differ. Broadly speaking the visual pattern appears most consistent
between the left three plots, and the two global estimates for the
Theil-Sen and plug-in statistics both appear as defensible estimates of
the relationship present in the bulk of the data.

\begin{figure}

\centering{

\pandocbounded{\includegraphics[keepaspectratio]{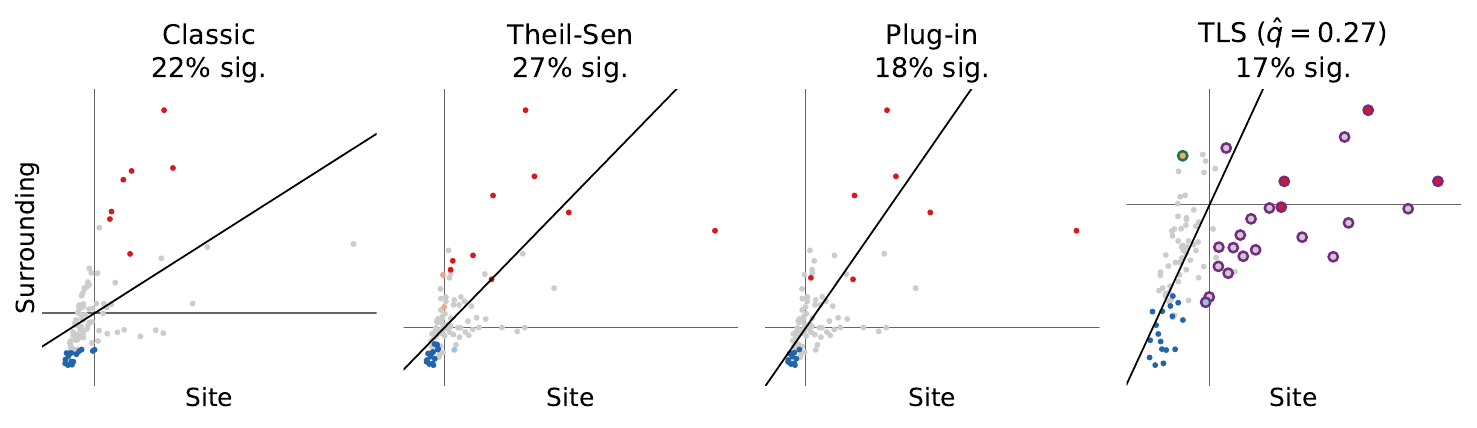}}

}

\caption{\label{fig-robust-scatterplot}Robust Moran scatterplots for
each estimator in the same realisation as
Figure~\ref{fig-single-clusters}. The axes are drawn with respect to
each estimators' location estimator: means for the classic Moran,
medians for the Theil-Sen and Plug-in estimator, and trimmed means for
the TLS estimator. Significant sites are coloured by quadrant (HH red,
LL blue, LH salmon, HL light blue); non-significant sites are grey. TLS
trimmed observations appear at their counterfactual lag position,
outlined by trim class (TrimHi purple, TrimLo green) and filled by their
Theil-Sen counterfactual class. Five observations in the TLS plot have
no neighbors, and must be omitted from the scatterplot because they have
no admissible lag.}

\end{figure}%

\subsection{Overall Simulation
Results}\label{overall-simulation-results}

We now present the results across the entire simulation frame, and
verify whether the expectations we outline in
Section~\ref{sec-expectations} hold.

\subsubsection{Cluster and outlier
classification}\label{cluster-and-outlier-classification}

\begin{figure}

\centering{

\pandocbounded{\includegraphics[keepaspectratio]{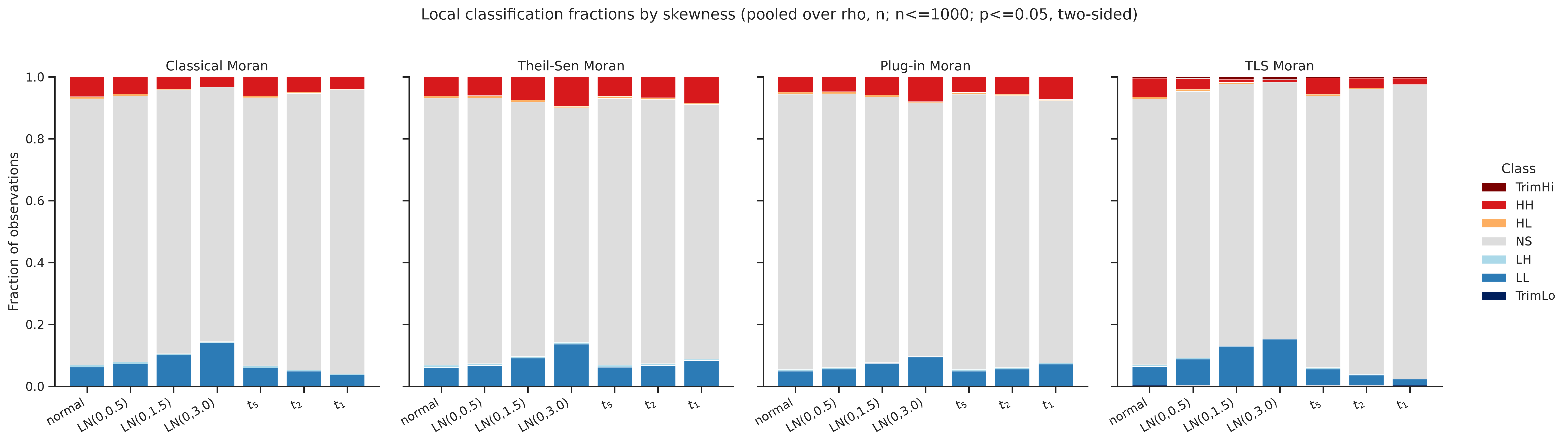}}

}

\caption{\label{fig-class-fractions}Local classification fractions by
skewness, pooled over \(\rho\) and \(n\) (\(n\leq1000\); \(p\leq0.05\),
two-sided).}

\end{figure}

Figure~\ref{fig-class-fractions} and Table~\ref{tbl-crosstabs} report
how each estimator assigns sites to the cluster/outlier quadrants
relative to the classical Moran statistic, pooled across \(\rho\) and
\(n\) (\(n\leq1000\)). We expect (H5) the classical Moran should label
increasingly many sites \(LL\) as \(\sigma\) grows in the \(SARLN\), and
we do see this in practice. Unfortunately, we do not see any estimator
manage to hold the classification distribution constant across skew and
kurtosis. All statistics reprise the same LL inflation as the global
statistic, as shown in Figure~\ref{fig-class-fractions}, although the
plug-in method seems to suppress this the most. The two also tend to get
dominated by the outliers as \(\nu\) decreases, with fewer and fewer
observations retaining their quadrant classifications. Interestingly,
the Theil-Sen and Plug-In statistics \emph{increase} significant
classifications as the tails get fatter; the TLS and classical Moran
statistics identify \emph{fewer} significant observations in the outlier
regime. This is only partially expected (H6).

\subsubsection{Estimator agreement}\label{estimator-agreement}

\begin{figure}

\begin{minipage}[t]{\linewidth}

\centering{

\pandocbounded{\includegraphics[keepaspectratio]{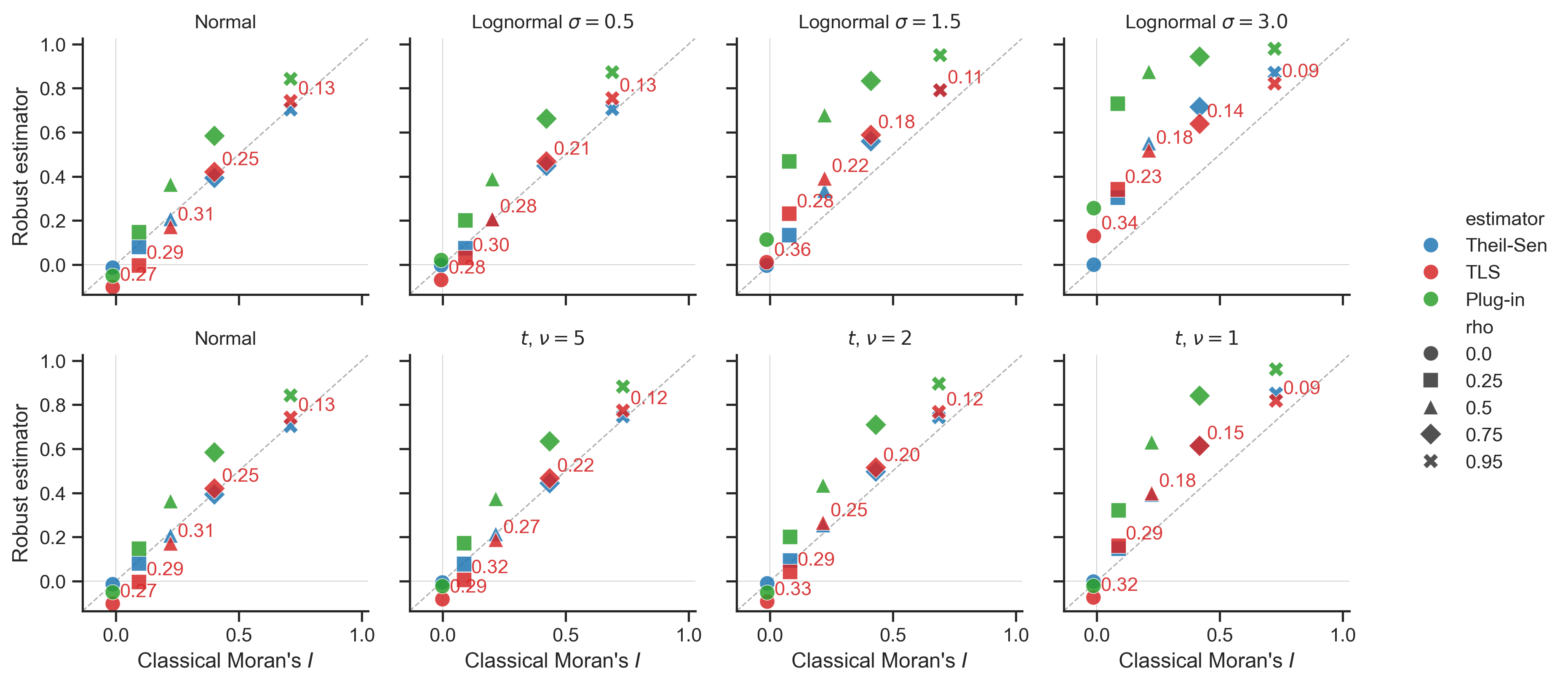}}

}

\subcaption{\label{fig-gscatter-100}\(n=100\)}

\end{minipage}%
\newline
\begin{minipage}[t]{\linewidth}

\centering{

\pandocbounded{\includegraphics[keepaspectratio]{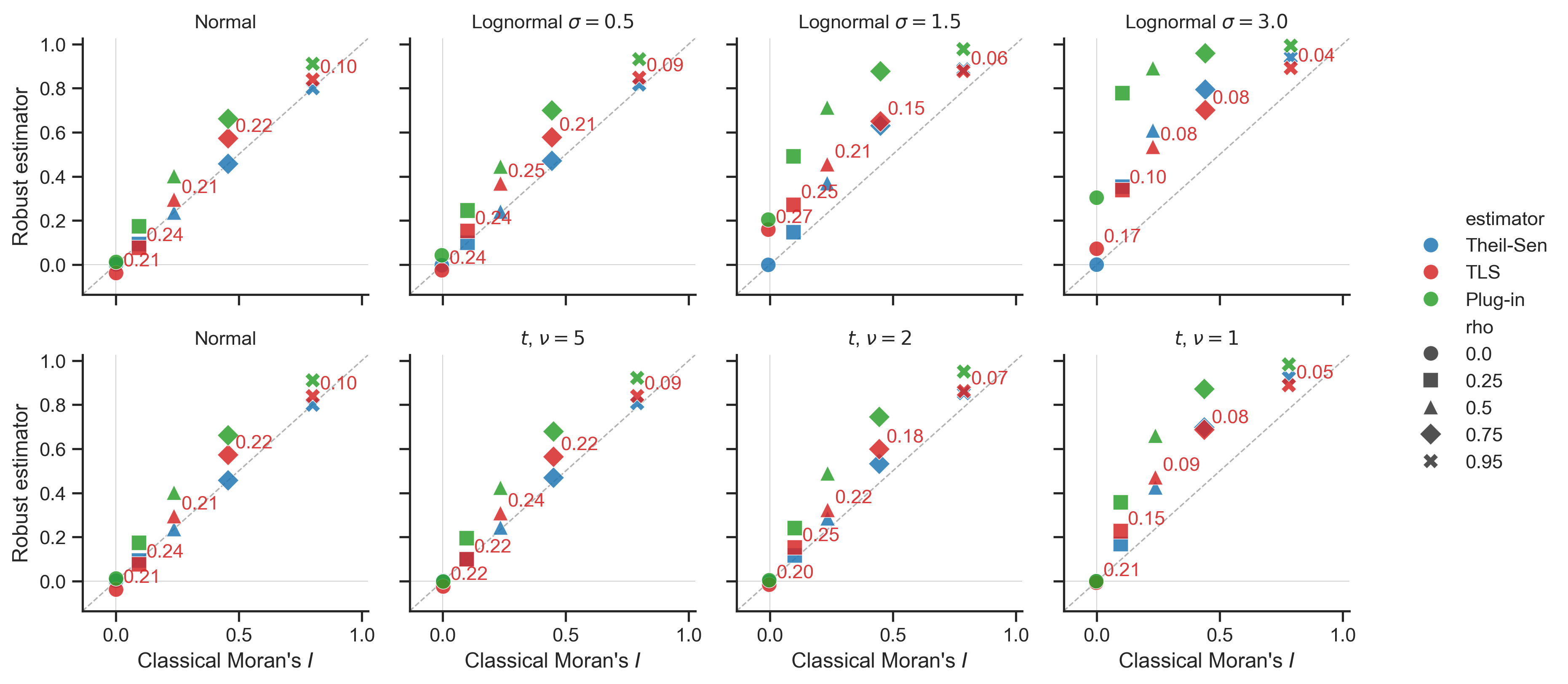}}

}

\subcaption{\label{fig-gscatter-1000}\(n=1000\)}

\end{minipage}%

\caption{\label{fig-global-scatter}Global robust estimates against
classical Moran's \(I\), faceted by simulation configuration; TLS points
annotated with the chosen \(\hat q\).}

\end{figure}%

Figure~\ref{fig-gscatter-100} and Figure~\ref{fig-gscatter-1000} show
the agreement for the average global estimate across replications for
each simulation configuration. The chosen TLS trimming fraction
\(\hat q\) is also annotated. We can see that all global estimators tend
to be larger than the classical Moran's \(I\) estimate when
\(\rho\neq0\). This is expected (H4), but the plug-in estimator is
\emph{always} largest for \(\rho\neq0\), which is not anticipated. It
seems that this estimator may inflate the estimate of spatial structure
present in data, given this structure. Beyond this for the global
statistics in Figure~\ref{fig-gscatter-100} and
Figure~\ref{fig-gscatter-1000}, the Theil-Sen and TLS statistics trade
off as closest to the global estimate (H1).

\begin{figure}

\centering{

\pandocbounded{\includegraphics[keepaspectratio]{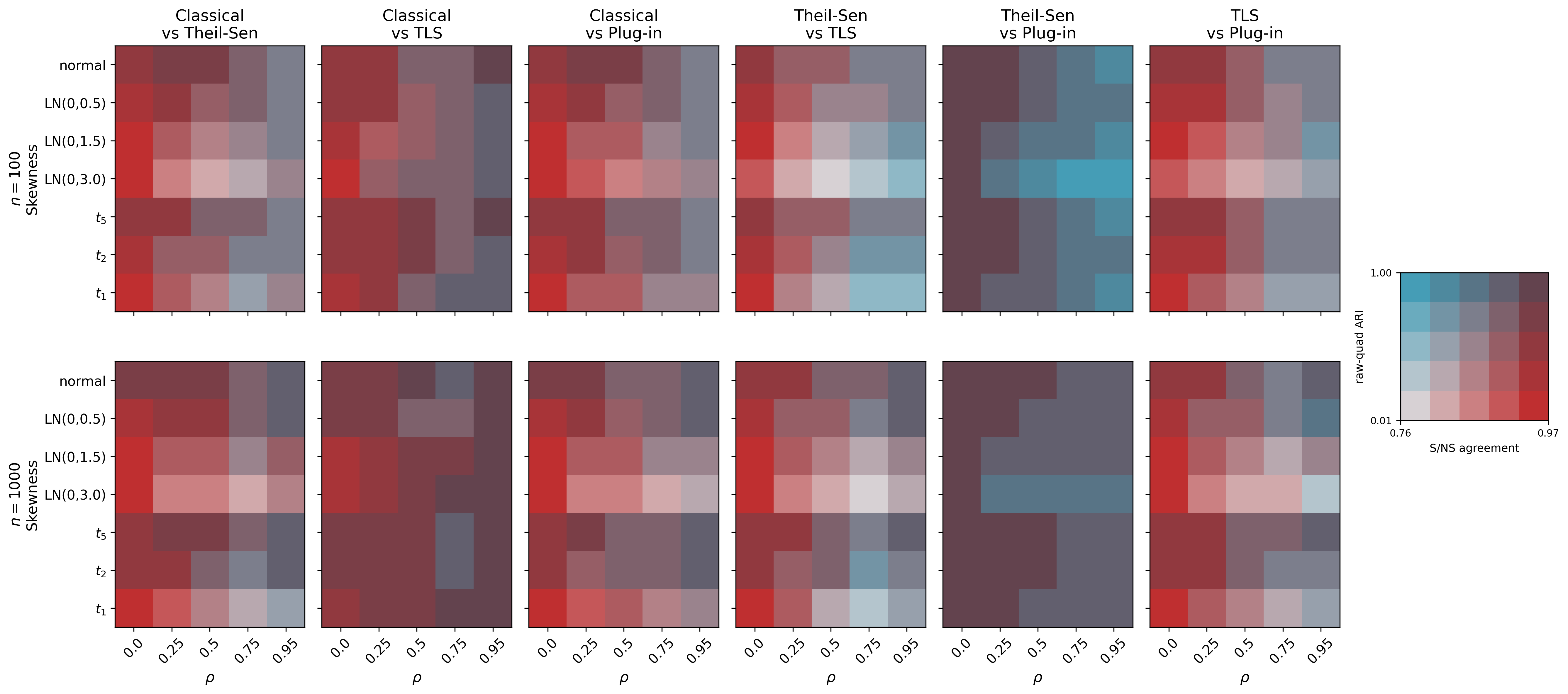}}

}

\caption{\label{fig-global-heatmap}Bivariate agreement between each pair
of local-Moran classifiers, decomposing agreement into two components:
the fraction of sites given the same significant/non-significant call
(\emph{S/NS agreement}, red axis) and the chance-corrected agreement
(adjusted Rand index) on the \(HH/LH/LL/HL\) quadrant with no
significance filter (\emph{raw-quadrant ARI}, blue axis). Red cells mark
pairs that agree on \emph{whether} a site carries signal but not on
\emph{which} quadrant; blue cells agree on quadrant but not
significance; dark cells agree on both; light cells on neither. Each
chroma axis is rescaled to its observed range and discretised into five
equal-interval bins (legend, lower right). Each cell pools all sites
across 100 reps; rows are skewness scenarios, columns are \(\rho\); top
block \(n=100\), bottom block \(n=1000\).}

\end{figure}%

In terms of the local statistics, Figure~\ref{fig-global-heatmap}
summarises the adjusted Rand index between local statistical estimators
for significance decisions and cluster/outlier classifications. There,
classifications strongly agree when the \(\rho\) is small, but get worse
as \(\rho\) increases. The plug-in and Theil-sen have higher general
agreement across all quadrant configurations, especially in the smaller
data, indicated by their stronger blue hues towards the right. The
classical and TLS statistics agree most strongly on the significance
decisions (and also quadrant decisions), given their generally red or
purple hues. Typical trimming fraction does not clearly seem to
correspond to experimental configuration: values tend to he higher for
small \(\rho\), but not for larger \(\sigma\) or smaller \(\nu\) (H9).

\subsubsection{Test size}\label{test-size}

\begin{figure}

\begin{minipage}[t]{\linewidth}

\centering{

\pandocbounded{\includegraphics[keepaspectratio]{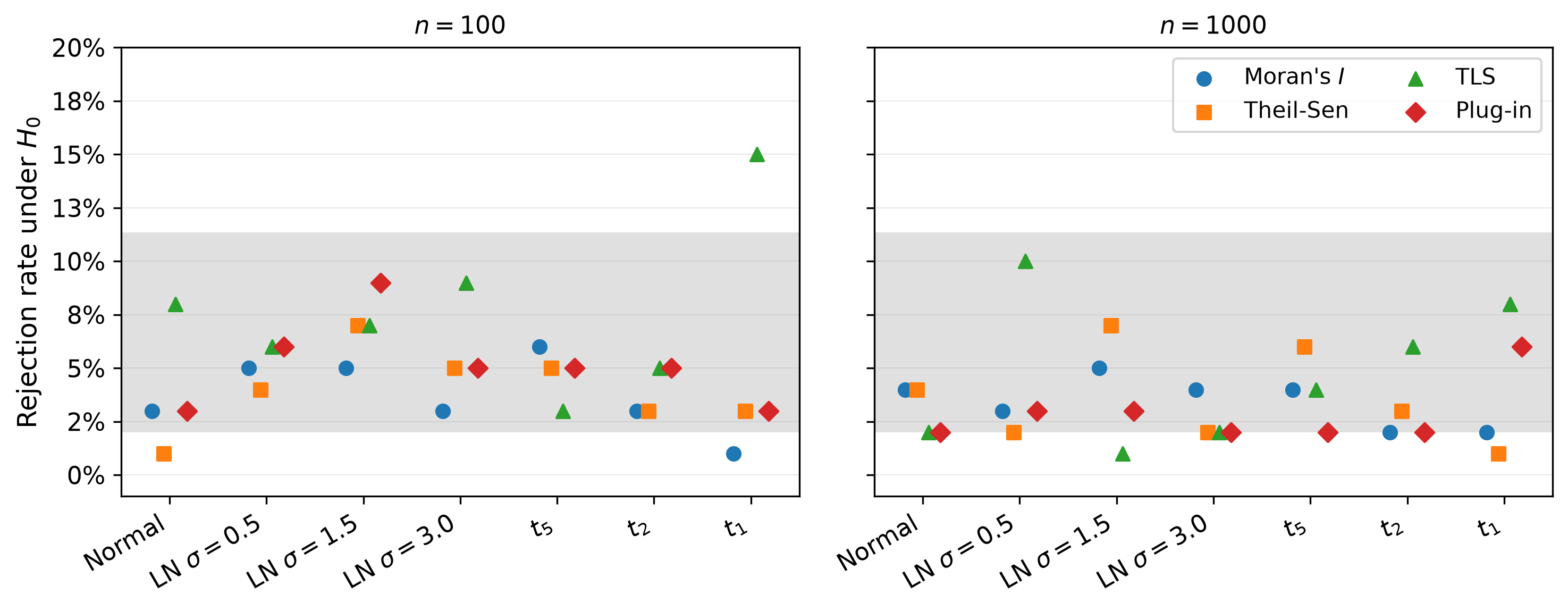}}

}

\subcaption{\label{fig-size-global}Global statistic}

\end{minipage}%
\newline
\begin{minipage}[t]{\linewidth}

\centering{

\pandocbounded{\includegraphics[keepaspectratio]{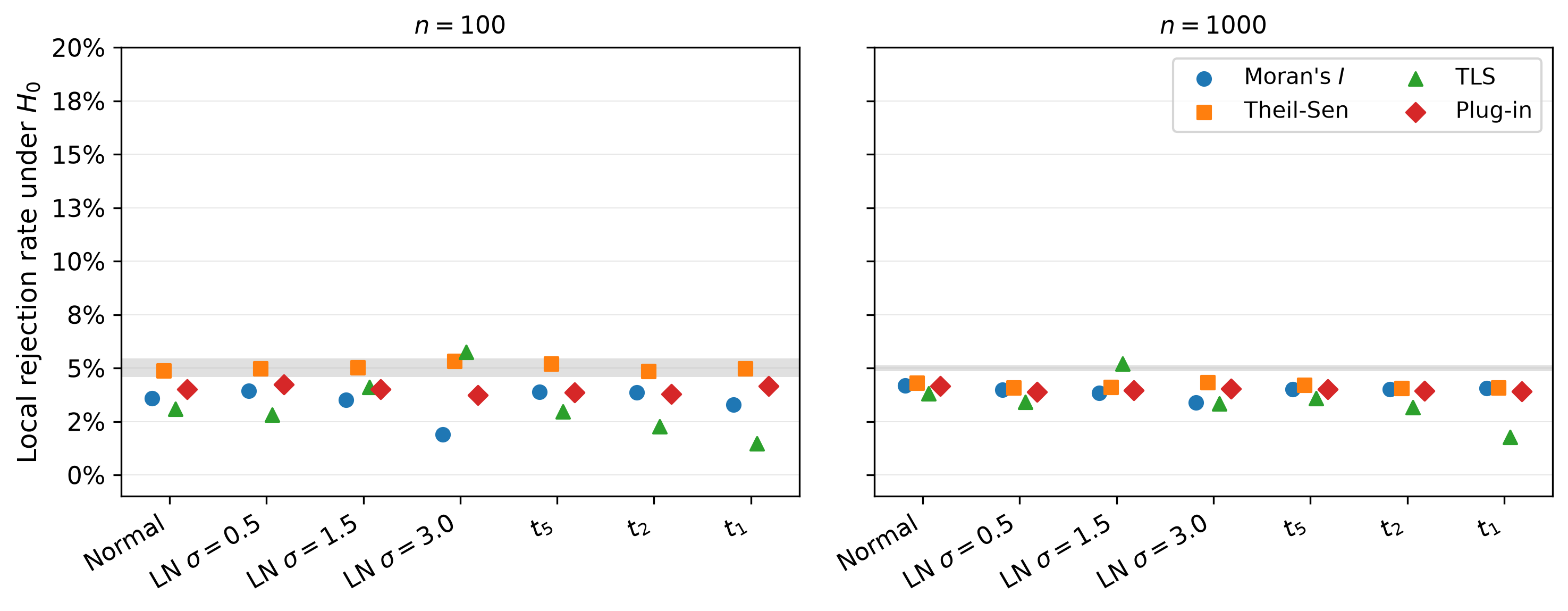}}

}

\subcaption{\label{fig-size-local}Local statistic}

\end{minipage}%

\caption{\label{fig-size}Rejection rate under \(H_0\) (\(\rho=0\)), 100
reps, \(\alpha=0.05\), pooled over replications among \(N=100\) and
\(N=1000\). The shaded band is the range of rejection rates consistent
with a correctly-sized test at \(\alpha=0.05\), obtained by inverting
the exact binomial test at the null-expected count (global: out of 100
replications; local: out of \(N\times100\) site-level tests pooled
across replications). This interval ignores spatial dependence among
local tests and is therefore narrower than the true Monte Carlo
variability.}

\end{figure}

\begin{figure}

\centering{

\pandocbounded{\includegraphics[keepaspectratio]{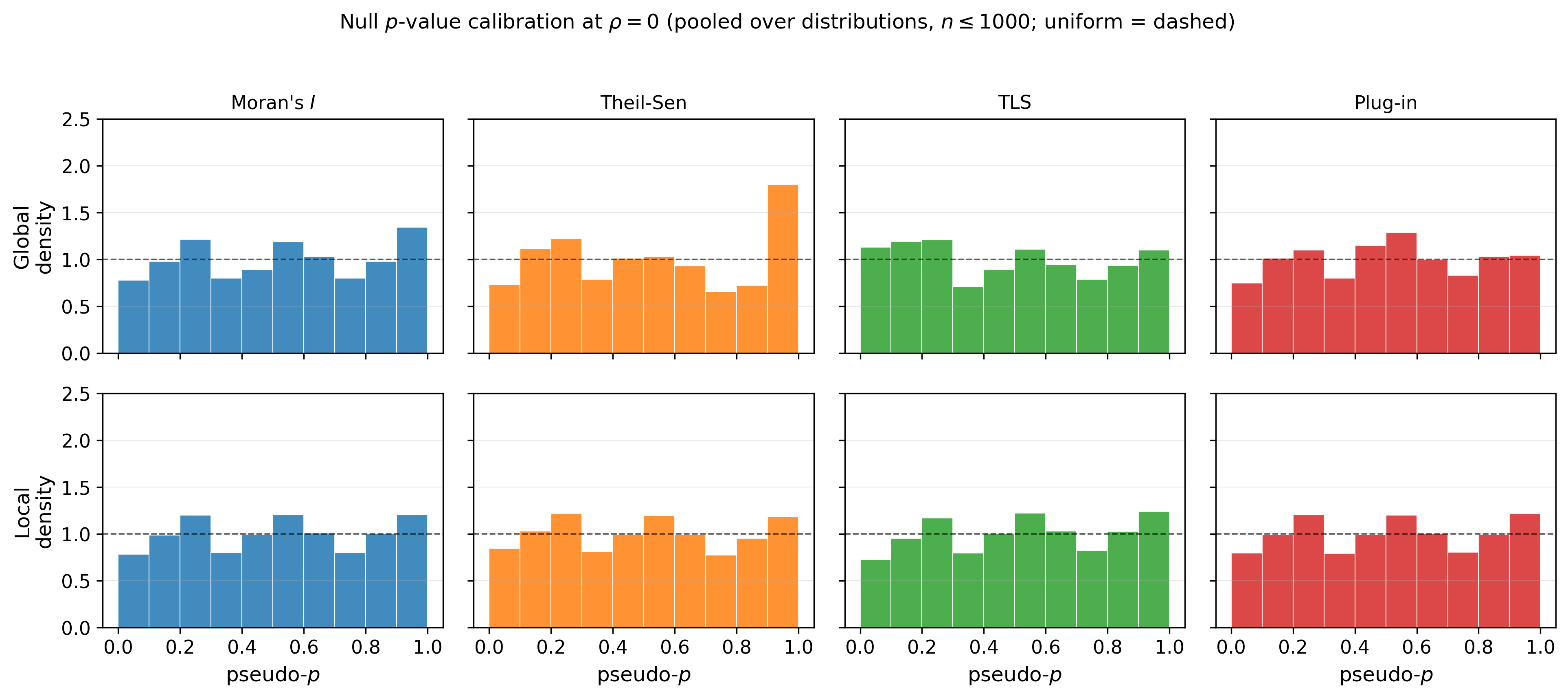}}

}

\caption{\label{fig-uniformity}Null \(p\)-value calibration at
\(\rho=0\), pooled across all distributions at \(n\leq1000\). Top row:
global pseudo-\(p\); bottom row: local pseudo-\(p\); one column per
estimator. A well-calibrated test is \(\mathrm{Uniform}(0,1)\),
i.e.~flat at the dashed reference line.}

\end{figure}

At \(\rho=0\) no spatial trend exists, so rejection rates should sit
near the nominal \(\alpha=0.05\) and null \(p\)-values should be uniform
(H2). Figure~\ref{fig-size} reports global and local rejection rates
across the distributional grid, and Figure~\ref{fig-uniformity} shows a
histogram of the p-values under the null, pooled across all simulations.
Under skew, global TLS and plug-in estimators are slightly over-sized,
as is the Theil-Sen regressor over moderate skew. However, most
worringly, the TLS regression has significant size issues at a few
different simulation configurations. This suggests that trimming is too
dramatic, and results in seeing spurious spatial patterns in the null
after cutting off what may not necessarily be true distributional
outliers. Over most of the simulation range, however, the Theil-Sen has
great size properties, and maps most closely to the classical Moran
statistic for both global and local statistics.

\subsubsection{Power}\label{power}

\begin{figure}

\begin{minipage}[t]{\linewidth}

\centering{

\pandocbounded{\includegraphics[keepaspectratio]{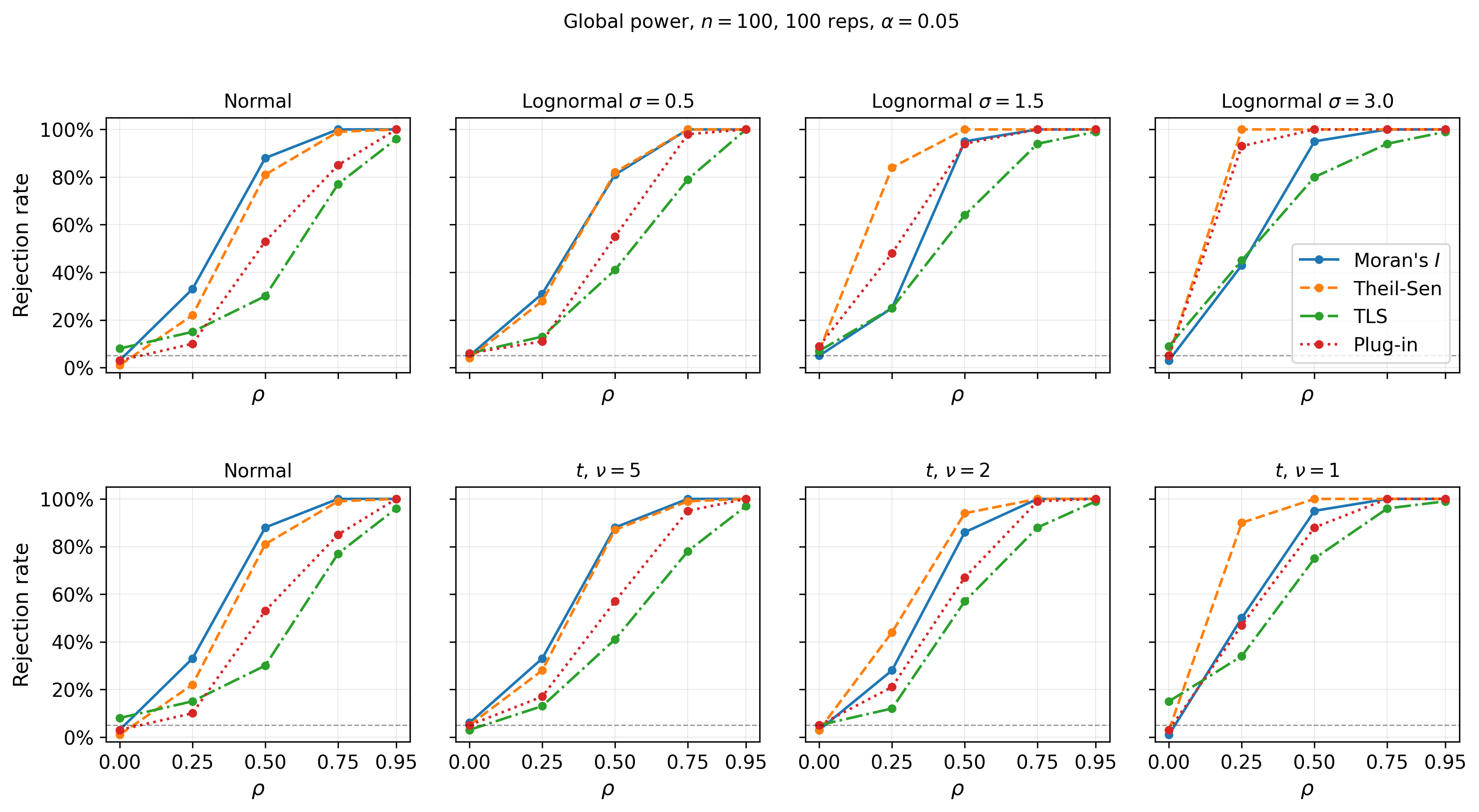}}

}

\subcaption{\label{fig-gpow-100}\(n=100\)}

\end{minipage}%
\newline
\begin{minipage}[t]{\linewidth}

\centering{

\pandocbounded{\includegraphics[keepaspectratio]{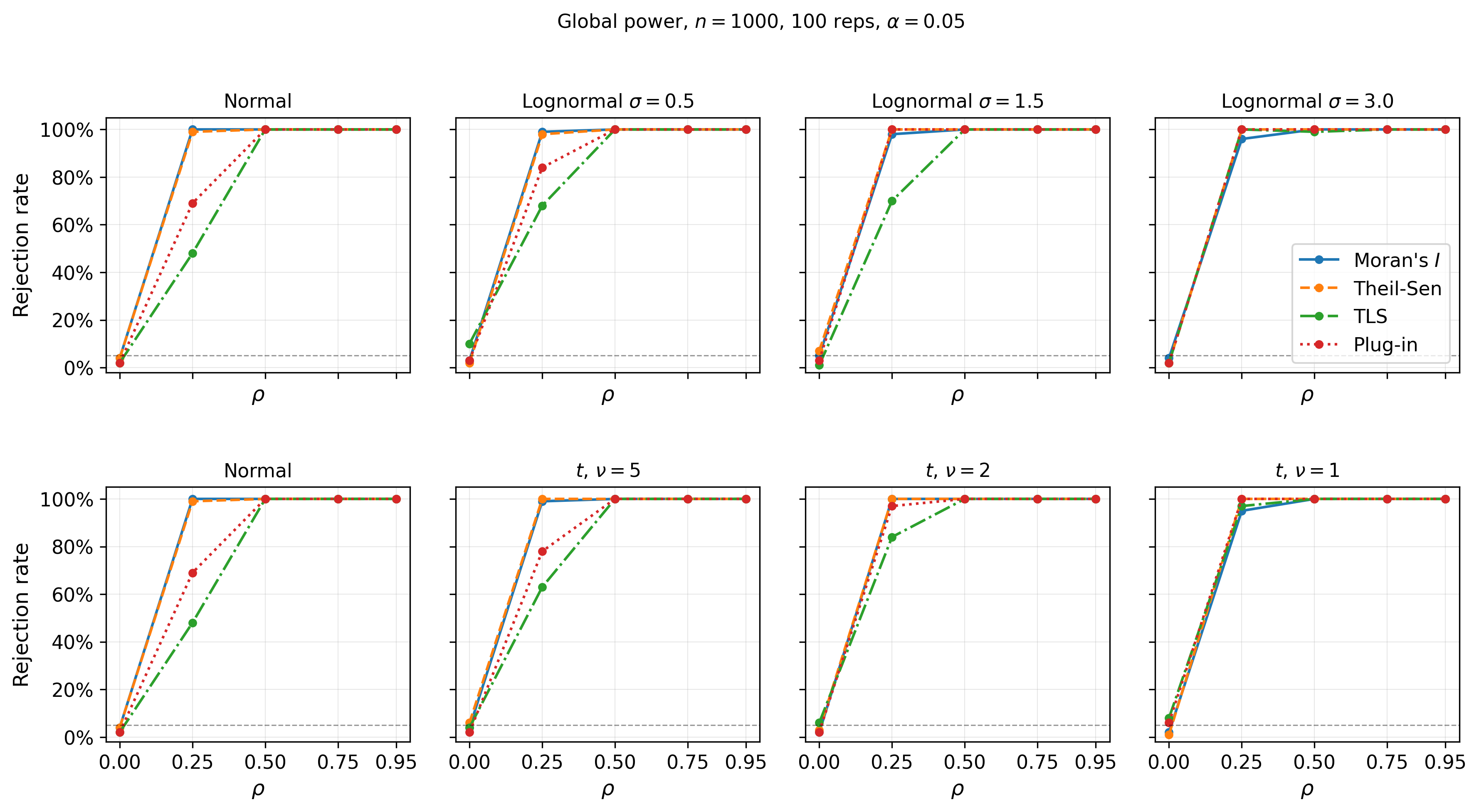}}

}

\subcaption{\label{fig-gpow-1000}\(n=1000\)}

\end{minipage}%

\caption{\label{fig-global-power}Global rejection rate across \(\rho\),
100 reps, \(\alpha=0.05\).}

\end{figure}

\begin{figure}

\begin{minipage}[t]{\linewidth}

\centering{

\pandocbounded{\includegraphics[keepaspectratio]{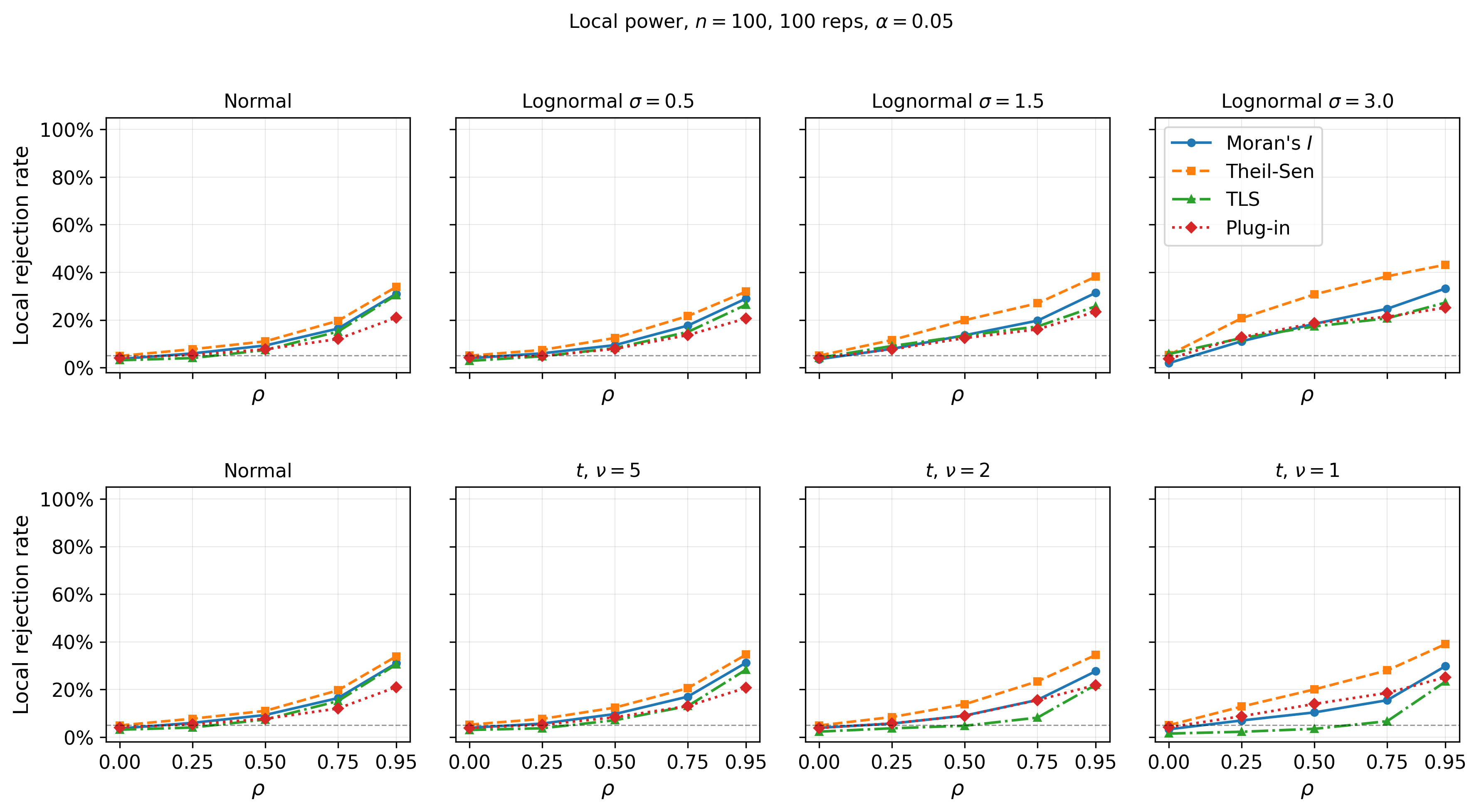}}

}

\subcaption{\label{fig-lpow-100}\(n=100\)}

\end{minipage}%
\newline
\begin{minipage}[t]{\linewidth}

\centering{

\pandocbounded{\includegraphics[keepaspectratio]{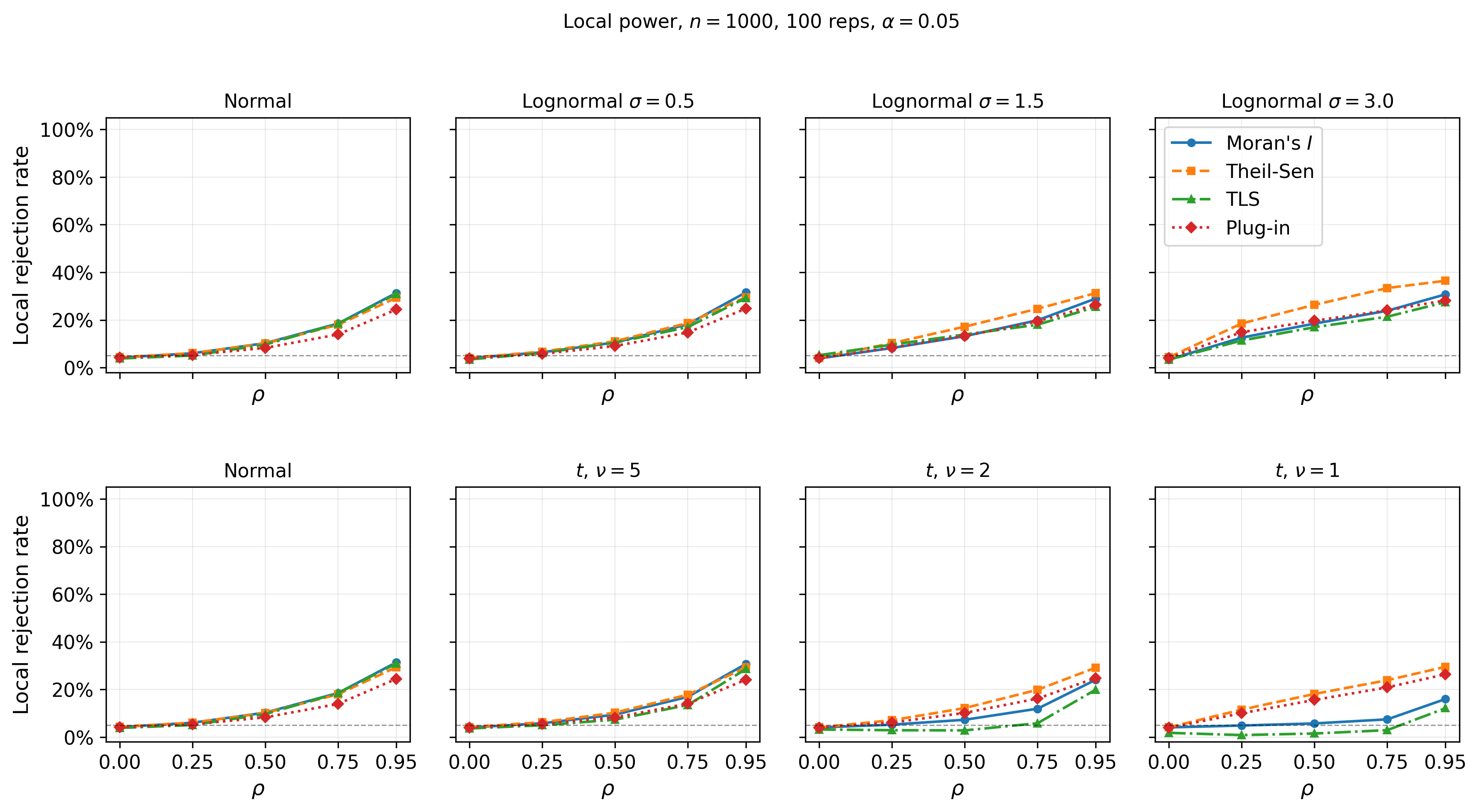}}

}

\subcaption{\label{fig-lpow-1000}\(n=1000\)}

\end{minipage}%

\caption{\label{fig-local-power}Local rejection rate across \(\rho\),
100 reps, \(\alpha=0.05\).}

\end{figure}

For \(\rho>0\) a trend exists, and power is the rejection rate (H3);
robustness to skew and kurtosis means power that does not collapse as
\(\sigma\) grows or \(\nu\) shrinks (H4). Figure~\ref{fig-global-power}
and Figure~\ref{fig-local-power} trace power across \(\rho\) for each
distribution at \(n=100\) and \(n=1000\). From these plots, we can see
that the Theil-Sen estimator is the \emph{dominating global robust
estimator}. It has the highest power at any \(\rho^+\) and acceptable
size under the null. This effect is clearest in the small sample. The
TLS estimator has a clear size problem, and over-rejects under the null.
The global plug-in estimator also behaves well, but is underpowered,
especially in small samples. A similar story holds for the local
statistics in Figure~\ref{fig-local-power}, where the TLS estimator has
severe lack of power in heavy-tailed situations (\(\nu=1\)). This is
likely because of the trimming \emph{removing} test hits that other
statistics detect as spatial outliers. Again, the Theil-Sen estimator
appears to be the most effective global \emph{and} local estimator.

\subsubsection{Counterfactual vs.~repeated C-step TLS
inference}\label{counterfactual-vs.-repeated-c-step-tls-inference}

We expect (H8) that the fast counterfactual TLS inference yields
\(p\)-values exceptionally close to the repeated C-step procedure. The
correlation of the two local statistics is 1 across all realisations,
and the two approaches agree on null rejection decisions \emph{at worst}
94.6\% of the time (\(n=100\) with high skew). Thus, if warranted, the
fast TLS estimator with counterfactual inference can be used instead of
a full rerun of the TLS algorithm at every local permutation
(Table~\ref{tbl-tls-cf-agree}).

\subsubsection{Computational
performance}\label{computational-performance}

We expect (H7) that Theil-Sen will be the slowest estimator, followed by
TLS, then plug-in and classical Moran, which should be substantially the
same timing.

\begin{figure}

\centering{

\pandocbounded{\includegraphics[keepaspectratio]{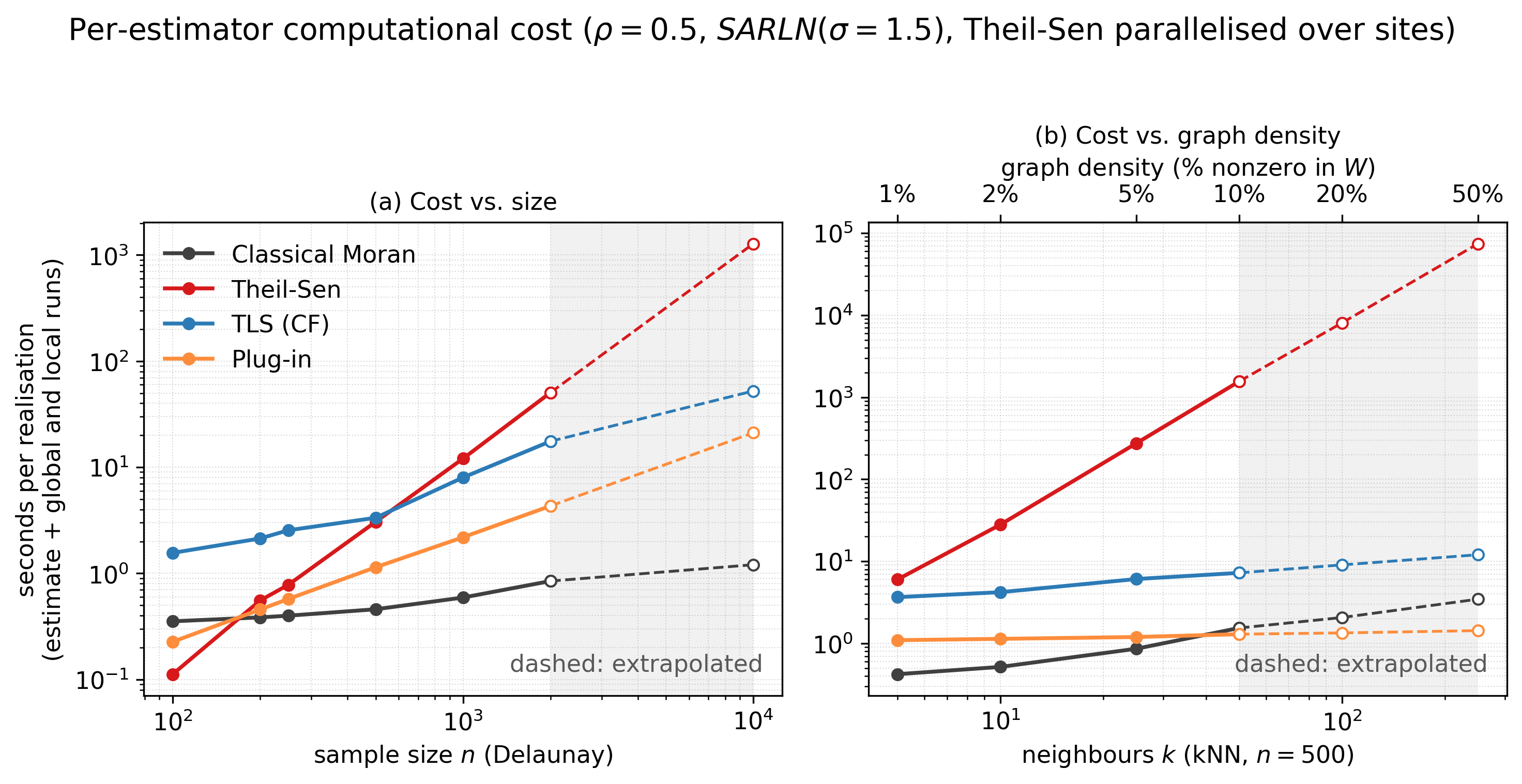}}

}

\caption{\label{fig-timing}Per-estimator computational cost
(\(\rho=0.5\), \(SARLN(\sigma=1.5)\)), as total wall-clock seconds per
realisation (estimate plus global and local permutation runs). (a) Cost
against sample size \(n\) on a Delaunay graph. (b) Cost against
neighbour count \(k\) on a \(k\)-NN graph at \(n=500\); the upper axis
shows the implied graph density (\(k/n\)). The \(k\in\{50,100,250\}\)
points (dashed, shaded) are projected from the fitted power law rather
than measured. Both axes are logarithmic.}

\end{figure}

Figure~\ref{fig-timing} reports the per-realisation wall-clock cost
(estimate plus global and local permutation runs) against sample size
(Delaunay) and graph density (kNN, \(n=500\)), at \(\rho=0.5\) under
\(SARLN(\sigma=1.5)\). The Theil-Sen conditional local permutation is
parallelised across sites using 8 threads; all other estimators run
single-threaded. The ordering matches H7---Theil-Sen is by far the most
expensive, TLS is second, and the plug-in and classical Moran statistics
are both cheap---but the gap is severe rather than marginal. At
\(n=1000\) the classical statistic costs \(0.6\)s, p;lug-in \(2.2\)s and
TLS \(7.7\)s, while Theil-Sen costs \(38\)s, \(65\times\) the classical
statistic. The conditional local permutation drives this: its cost grows
as \(O(\text{reps}\cdot\text{perms}\cdot n\cdot k^2)\), and the measured
density exponent (\(k^{2.38}\) over \(k\in\{5,10,25\}\)) recovers the
quadratic-in-\(k\) term. A literal measurement at \(k\in\{50,100,250\}\)
would take roughly \(23\)m, \(2\)h, and \(18\)h per realisation
respectively, so those three density points are reported via the fitted
power law and drawn as the dashed, shaded segment in panel (b).

\section{Discussion and Conclusion}\label{discussion-and-conclusion}

Moran-family measures of local association are useful for characterising
whether observations are spatial outliers---strikingly different values
when compared to other values nearby---or spatial clusters, composed of
similar values all on the same side of the distribution. They are doubly
useful because they offer a simple visualization, the Moran Scatterplot,
which graphically represents their relation to the global statistic,
which measures the spatial association across observations in the whole
map. Unfortunately, the classic Moran statistic is sensitive to
\emph{distributional outliers}, values which are unusual regardless of
where they are located in the map. This means that the classic statistic
used to detect \emph{spatial outliers} is substantially misled by
outliers which are unusual regardless of their location. We argue that
this is quite problematic for one of the most commonly-used exploratory
spatial data analysis statistics.

To mitigate this issue, we cover a variety of robust estimation
techniques for Moran-style local and global measures of spatial
association. We find (H1) that the Theil-Sen and TLS global estimator
trade off as most similar to the classical global estimate over
simulations, while the global plug-in estimator significantly
over-states the magnitude of association. (H2) Unexpectedly, TLS
estimators tend to be too large under the null, while TLS statistics and
plug-in estimators tend to be more appropriately sized, however
\(p\)-values are practically uniform for all estimators. (H3) As
expected, the Theil-Sen estimator is indeed the most powerful statistic
across all cases, but the plug-in estimators obtain similar power in
larger data. (H4) We do find evidence that the Moran statistic is
understated globally under conditions of high skew, and proportions of
opposite-tail outliers tend to increase (H5). We find that the (H6)
Theil-Sen and plug-in statistics agree significantly on the local
estimates, and find that the TLS local statistics are most similar to
the classical Moran statistics. In terms of speed, (H7) Theil-sen is
unexpectedly fast in small data, but our anticipated speed ranking is
recovered by \(n=1000\). Despite the serious drawbacks identified for
the TLS estimator, we do find that (H8) the counterfactual inference
strategy yields substantially the same results as the significantly more
expensive repeated-survivor inference strategy, while trim fraction (H9)
only seems to decline clearly with spatial patterning, not tail
thickness or skew strength.

Synthesizing these findings overall, we suggest that the Theil-Sen
estimator is powerful, efficient, and fast enough for common use in
moderately-sized samples. Given the computational intensity for this
technique, we think that the TLS estimator is unlikely to yield useful
or tractable estimators going forward. The TLS estimator has significant
size and power issues in precisely the situations it must be most
useful: high skew, fat distributional tails, as well as under no
patterning. Beyond this, the robust plug-in estimator of
\citet{arbia2025robust} also achieves reasonable power and robustness to
skew and distributional outliers in large data. Fortunately, the plug-in
and Theil-Sen estimators share the same positions and axes within the
Robust Moran Scatterplot and ultimately provide very similar local
estimates. The two do differ on the strength of global patterning,
however, with the plug-in estimator seeming to over-state the magnitude
of map patterning. Between these two estimators, the Theil-Sen estimator
should be used if the computational cost can be paid, since it has the
the best combination of power, size, and sensitivity to spatial
outliers; if it cannot, the plug-in estimators can be used. More
generally, these should be provide new ``default'' estimators for
generating local indicators of spatial association.

\section{Appendix}\label{appendix}

{

\begin{longtable}[]{@{}
  >{\raggedright\arraybackslash}p{(\linewidth - 16\tabcolsep) * \real{0.1500}}
  >{\raggedright\arraybackslash}p{(\linewidth - 16\tabcolsep) * \real{0.1625}}
  >{\raggedright\arraybackslash}p{(\linewidth - 16\tabcolsep) * \real{0.1375}}
  >{\raggedright\arraybackslash}p{(\linewidth - 16\tabcolsep) * \real{0.0750}}
  >{\raggedright\arraybackslash}p{(\linewidth - 16\tabcolsep) * \real{0.0750}}
  >{\raggedright\arraybackslash}p{(\linewidth - 16\tabcolsep) * \real{0.0750}}
  >{\raggedright\arraybackslash}p{(\linewidth - 16\tabcolsep) * \real{0.0750}}
  >{\raggedright\arraybackslash}p{(\linewidth - 16\tabcolsep) * \real{0.1250}}
  >{\raggedright\arraybackslash}p{(\linewidth - 16\tabcolsep) * \real{0.1250}}@{}}

\caption{\label{tbl-crosstabs}Quadrant crosstabs: classical Moran
quadrant (rows) against each robust estimator's assignment,
row-normalised so each classical quadrant sums to 100\% across the
robust labels. By skewness, pooled over \(\rho\) and \(n\)
(\(n\leq1000\)). \texttt{TrimHi}/\texttt{TrimLo} give the share of those
sites the TLS removes as distributional outliers.}

\tabularnewline

\toprule\noalign{}
\begin{minipage}[b]{\linewidth}\raggedright
skewness
\end{minipage} & \begin{minipage}[b]{\linewidth}\raggedright
estimator
\end{minipage} & \begin{minipage}[b]{\linewidth}\raggedright
classic
\end{minipage} & \begin{minipage}[b]{\linewidth}\raggedright
HH
\end{minipage} & \begin{minipage}[b]{\linewidth}\raggedright
LH
\end{minipage} & \begin{minipage}[b]{\linewidth}\raggedright
LL
\end{minipage} & \begin{minipage}[b]{\linewidth}\raggedright
HL
\end{minipage} & \begin{minipage}[b]{\linewidth}\raggedright
TrimHi
\end{minipage} & \begin{minipage}[b]{\linewidth}\raggedright
TrimLo
\end{minipage} \\
\midrule\noalign{}
\endhead
\bottomrule\noalign{}
\endlastfoot
normal & Theil-Sen & HH & 90\% & 1\% & 0\% & 9\% & & \\
& & LH & 1\% & 82\% & 16\% & 0\% & & \\
& & LL & 0\% & 9\% & 90\% & 1\% & & \\
& & HL & 16\% & 0\% & 1\% & 82\% & & \\
& --- & & & & & & & \\
& TLS & HH & 93\% & 0\% & 0\% & 7\% & 1\% & 0\% \\
& & LH & 0\% & 87\% & 12\% & 0\% & 0\% & 1\% \\
& & LL & 0\% & 6\% & 93\% & 0\% & 0\% & 1\% \\
& & HL & 12\% & 0\% & 0\% & 87\% & 1\% & 0\% \\
& --- & & & & & & & \\
& Plug-in & HH & 90\% & 1\% & 0\% & 9\% & & \\
& & LH & 1\% & 82\% & 16\% & 0\% & & \\
& & LL & 0\% & 9\% & 90\% & 1\% & & \\
& & HL & 16\% & 0\% & 1\% & 82\% & & \\
--- & & & & & & & & \\
LN(0,0.5) & Theil-Sen & HH & 92\% & 0\% & 0\% & 8\% & & \\
& & LH & 18\% & 66\% & 13\% & 3\% & & \\
& & LL & 2\% & 12\% & 78\% & 8\% & & \\
& & HL & 24\% & 0\% & 0\% & 76\% & & \\
& --- & & & & & & & \\
& TLS & HH & 90\% & 0\% & 0\% & 9\% & 1\% & 0\% \\
& & LH & 0\% & 81\% & 18\% & 0\% & 0\% & 1\% \\
& & LL & 0\% & 10\% & 89\% & 0\% & 0\% & 0\% \\
& & HL & 21\% & 0\% & 0\% & 78\% & 1\% & 0\% \\
& --- & & & & & & & \\
& Plug-in & HH & 92\% & 0\% & 0\% & 8\% & & \\
& & LH & 18\% & 66\% & 13\% & 3\% & & \\
& & LL & 2\% & 12\% & 78\% & 8\% & & \\
& & HL & 24\% & 0\% & 0\% & 76\% & & \\
--- & & & & & & & & \\
LN(0,1.5) & Theil-Sen & HH & 96\% & 0\% & 0\% & 4\% & & \\
& & LH & 52\% & 32\% & 9\% & 7\% & & \\
& & LL & 9\% & 16\% & 63\% & 12\% & & \\
& & HL & 46\% & 0\% & 0\% & 54\% & & \\
& --- & & & & & & & \\
& TLS & HH & 82\% & 0\% & 0\% & 15\% & 2\% & 0\% \\
& & LH & 0\% & 52\% & 47\% & 0\% & 1\% & 1\% \\
& & LL & 0\% & 13\% & 87\% & 0\% & 0\% & 0\% \\
& & HL & 31\% & 0\% & 0\% & 68\% & 1\% & 0\% \\
& --- & & & & & & & \\
& Plug-in & HH & 96\% & 0\% & 0\% & 4\% & & \\
& & LH & 52\% & 32\% & 9\% & 7\% & & \\
& & LL & 9\% & 16\% & 63\% & 12\% & & \\
& & HL & 46\% & 0\% & 0\% & 54\% & & \\
--- & & & & & & & & \\
LN(0,3.0) & Theil-Sen & HH & 99\% & 0\% & 0\% & 1\% & & \\
& & LH & 75\% & 14\% & 5\% & 6\% & & \\
& & LL & 24\% & 12\% & 52\% & 11\% & & \\
& & HL & 72\% & 0\% & 0\% & 28\% & & \\
& --- & & & & & & & \\
& TLS & HH & 84\% & 0\% & 0\% & 11\% & 5\% & 0\% \\
& & LH & 0\% & 45\% & 54\% & 0\% & 1\% & 0\% \\
& & LL & 0\% & 8\% & 92\% & 0\% & 0\% & 0\% \\
& & HL & 21\% & 0\% & 0\% & 78\% & 1\% & 0\% \\
& --- & & & & & & & \\
& Plug-in & HH & 99\% & 0\% & 0\% & 1\% & & \\
& & LH & 75\% & 14\% & 5\% & 6\% & & \\
& & LL & 24\% & 12\% & 52\% & 11\% & & \\
& & HL & 72\% & 0\% & 0\% & 28\% & & \\
--- & & & & & & & & \\
\(t_{5}\) & Theil-Sen & HH & 89\% & 1\% & 0\% & 10\% & & \\
& & LH & 2\% & 80\% & 18\% & 0\% & & \\
& & LL & 0\% & 10\% & 89\% & 1\% & & \\
& & HL & 18\% & 0\% & 1\% & 80\% & & \\
& --- & & & & & & & \\
& TLS & HH & 93\% & 0\% & 0\% & 7\% & 1\% & 0\% \\
& & LH & 0\% & 86\% & 14\% & 0\% & 0\% & 1\% \\
& & LL & 0\% & 7\% & 93\% & 0\% & 0\% & 1\% \\
& & HL & 14\% & 0\% & 0\% & 86\% & 1\% & 0\% \\
& --- & & & & & & & \\
& Plug-in & HH & 89\% & 1\% & 0\% & 10\% & & \\
& & LH & 2\% & 80\% & 18\% & 0\% & & \\
& & LL & 0\% & 10\% & 89\% & 1\% & & \\
& & HL & 18\% & 0\% & 1\% & 80\% & & \\
--- & & & & & & & & \\
\(t_{2}\) & Theil-Sen & HH & 87\% & 2\% & 1\% & 10\% & & \\
& & LH & 4\% & 74\% & 21\% & 1\% & & \\
& & LL & 1\% & 10\% & 87\% & 2\% & & \\
& & HL & 21\% & 1\% & 4\% & 75\% & & \\
& --- & & & & & & & \\
& TLS & HH & 91\% & 0\% & 0\% & 8\% & 1\% & 0\% \\
& & LH & 0\% & 82\% & 17\% & 0\% & 0\% & 1\% \\
& & LL & 0\% & 8\% & 91\% & 0\% & 0\% & 1\% \\
& & HL & 16\% & 0\% & 0\% & 83\% & 1\% & 0\% \\
& --- & & & & & & & \\
& Plug-in & HH & 87\% & 2\% & 1\% & 10\% & & \\
& & LH & 4\% & 74\% & 21\% & 1\% & & \\
& & LL & 1\% & 10\% & 87\% & 2\% & & \\
& & HL & 21\% & 1\% & 4\% & 75\% & & \\
--- & & & & & & & & \\
\(t_{1}\) & Theil-Sen & HH & 67\% & 8\% & 13\% & 11\% & & \\
& & LH & 20\% & 50\% & 27\% & 3\% & & \\
& & LL & 14\% & 11\% & 67\% & 8\% & & \\
& & HL & 29\% & 2\% & 18\% & 50\% & & \\
& --- & & & & & & & \\
& TLS & HH & 87\% & 0\% & 0\% & 13\% & 1\% & 0\% \\
& & LH & 0\% & 76\% & 23\% & 0\% & 0\% & 1\% \\
& & LL & 0\% & 12\% & 87\% & 0\% & 0\% & 1\% \\
& & HL & 23\% & 0\% & 0\% & 76\% & 1\% & 0\% \\
& --- & & & & & & & \\
& Plug-in & HH & 67\% & 8\% & 13\% & 11\% & & \\
& & LH & 20\% & 50\% & 27\% & 3\% & & \\
& & LL & 14\% & 11\% & 67\% & 8\% & & \\
& & HL & 29\% & 2\% & 18\% & 50\% & & \\

\end{longtable}

}

{

\begin{longtable}[]{@{}lrrrl@{}}

\caption{\label{tbl-tls-cf-agree}Agreement between counterfactual and
full C-step TLS local inference, by distribution and \(n\) (subset: all
sites). \(r_p\): Pearson correlation of pseudo-\(p\)-values; MAD(\(p\)):
mean absolute difference; agreement: fraction of sites with identical
rejection decisions at \(\alpha=0.05\).}

\tabularnewline

\toprule\noalign{}
distribution & \(n\) & \(r_p\) & MAD(\(p\)) & agreement \\
\midrule\noalign{}
\endhead
\bottomrule\noalign{}
\endlastfoot
LN(0,0.5) & 100 & 0.965 & 0.049 & 96.6\% \\
LN(0,1.5) & 100 & 0.928 & 0.068 & 95.1\% \\
LN(0,3.0) & 100 & 0.919 & 0.068 & 94.6\% \\
normal & 100 & 0.969 & 0.047 & 96.6\% \\
\(t_{1}\) & 100 & 0.966 & 0.048 & 97.8\% \\
\(t_{2}\) & 100 & 0.97 & 0.045 & 97.3\% \\
\(t_{5}\) & 100 & 0.97 & 0.046 & 96.8\% \\
LN(0,0.5) & 1000 & 0.974 & 0.044 & 96.0\% \\
LN(0,1.5) & 1000 & 0.897 & 0.082 & 95.4\% \\
LN(0,3.0) & 1000 & 0.942 & 0.055 & 97.4\% \\
normal & 1000 & 0.977 & 0.041 & 96.2\% \\
\(t_{1}\) & 1000 & 0.984 & 0.033 & 98.9\% \\
\(t_{2}\) & 1000 & 0.98 & 0.039 & 97.8\% \\
\(t_{5}\) & 1000 & 0.978 & 0.04 & 96.8\% \\

\end{longtable}

}

\bibliography{citations.bib}

\end{document}